\documentclass{article}
\usepackage[hyphens]{url}
\usepackage[utf8]{inputenc}
\usepackage{graphicx}
\usepackage{authblk}
\usepackage[english]{babel}
\usepackage{setspace}
\usepackage{fancyhdr}
\usepackage{array}
\usepackage[margin=1in]{geometry}
\usepackage{float}
\usepackage{amsmath}
\usepackage{amssymb}
\usepackage{bbm}
\usepackage{bm}
\usepackage{scalerel,stackengine}
\usepackage[final]{pdfpages}
\usepackage{longtable}
\usepackage{chngcntr}
\usepackage{placeins}
\usepackage{microtype}
\usepackage{tikz}
\usepackage{makecell}
\usepackage{hyperref}
\usepackage{subcaption}
\usepackage{rotating}
\usepackage{soul}
\usepackage{tabularx} 
\usepackage{pdfpages}
\usepackage[square,numbers]{natbib}
\usepackage{multirow}
\hypersetup{
    colorlinks = true,
    citecolor = {blue},
    urlcolor = {blue},
    menucolor = {blue},
    linkcolor = {blue}
}

\makeatletter
\patchcmd{\@maketitle}{\LARGE \@title}{\fontsize{18}{20}\selectfont\textbf{\@title}}{}{}
\makeatother

\title{On the inclusion of non-concurrent controls in platform trials with an interim analysis}

\author[1]{Pavla Krotka}
\author[2]{Martin Posch}
\author[1]{Marta Bofill Roig\thanks{marta.bofillroig@meduniwien.ac.at}}
\affil[1]{Department of Statistics and Operations Research and Institute for Research and Innovation in Health (IRIS), Universitat Politècnica de Catalunya - BarcelonaTech (UPC), Barcelona, Spain}
\affil[2]{Center for Medical Data Science, Medical University of Vienna, Vienna 1090, Austria}
\date{}         
\newcommand{\probP}{\text{I\kern-0.15em P}}

\begin{document}
	
	\maketitle

\begin{abstract}
The analysis of platform trials can be enhanced by utilizing non-concurrent controls. Since including this data might also introduce bias in the treatment effect estimators if time trends are present, methods for incorporating non-concurrent controls adjusting for time have been proposed. However, so far their behavior has not been systematically investigated in platform trials that include interim analyses. To evaluate the impact of an interim analysis in trials utilizing non-concurrent controls, we consider a platform trial featuring two experimental arms and a shared control, with the second experimental arm entering later. We focus on a frequentist regression model that uses non-concurrent controls to estimate the treatment effect of the second arm and adjusts for time using a step function to account for temporal changes. We show that performing an interim analysis in Arm 1 may introduce bias in the point estimation of the effect in Arm 2, if the regression model is used without adjustment, and investigate how the marginal bias and bias conditional on the first arm continuing after the interim depend on different trial design parameters. Moreover, we propose a new estimator of the treatment effect in Arm 2, aiming to eliminate the bias introduced by both the interim analysis in Arm 1 and the time trends, and evaluate its performance in a simulation study. The newly proposed estimator is shown to substantially reduce the bias and type I error rate inflation while leading to power gains compared to an analysis using only concurrent controls.
\end{abstract}

%%%%%%%%%%%%%%%%%%%%%%%%%%%%%%
%%%%%%%%%%%%%%%%%%%%%%%%%%%%%%%%%%%%%%  
\section{Introduction}\label{sec_introduction}

Recent years have seen an increased demand for innovative clinical trial designs to accelerate drug development and optimize the use of resources \cite{Meyer2020Evolution}. Platform trials provide a framework for investigating multiple treatment arms simultaneously, while allowing new promising arms that become available during the trial to enter later on \cite{Koenig2024Current, Woodcock2017Master}. Typically, all treatment arms are compared to a shared control arm, which leads to a reduction in the required sample size compared to conducting separate clinical trials. Moreover, interim analyses are often included to enable an early decision on the efficacy or futility of the treatment arms and further acceleration of the drug development process \cite{Angus2019Adaptive}. For late-entering arms, the shared control arm is split into two groups: the concurrent (CC) and non-concurrent controls (NCC), where the first group refers to patients randomized to the control arm while the evaluated arm was also open for randomization, and the latter to patients allocated to the control arm before the evaluated arm joined the platform. Over the last years, it has been critically discussed whether and how the non-concurrent controls should be used for the analysis of late-entering arms, since direct pooling of concurrent and non-concurrent control may lead to biased treatment effect estimates and type I error rate inflation due to temporal drifts \cite{Bofill2022Review, Dodd2021Platform}. Such temporal drifts may result from, for instance, changes in the patient population, standard of care treatment, or seasonal effects \cite{FDA2023Master}.

Several approaches aiming to utilize non-concurrent controls while ensuring valid statistical inference have recently been proposed. In particular, Lee and Wason \cite{Lee2020Including} and Bofill Roig et al. \cite{Bofill2022Model} considered incorporating the NCC data using a regression model with categorical time adjustment, including the periods in the trial as a fixed effect, where periods are defined as time intervals bounded by any experimental arm entering or leaving the platform. This approach was investigated in a simple platform trial with two treatment arms and it was shown that this model yields unbiased treatment effect estimates if the time trends are equal across all arms and additive on the model scale \cite{Bofill2022Model}. Krotka et al. \cite{krotka2025statistical} examined the performance of this model in more complex platform trials with an arbitrary number of treatment arms and extended the frequentist methodology for incorporating non-concurrent controls by proposing more flexible methods, such as spline regression or mixed models. Another frequentist method proposed by Marschner and Schou \cite{Marschner2022Analysis} uses a network meta-analysis approach to analyze the platform trial as a network of direct randomized comparisons and indirect non-randomized comparisons. Methods based on propensity score weighting for incorporating NCC in platform trials with time trends under the framework of causal inference were considered by Guo et al. \cite{guo2024treatment}. Among Bayesian approaches considered for utilizing NCC data are the Bayesian Time Machine \cite{Saville2022Bayesian} and the meta-analytic-predictive (MAP) prior approach \cite{Schmidli2014Robust, Weber2021Applying}. The Time Machine approach uses a Bayesian hierarchical model to smooth the control response over calendar time intervals. The MAP prior approach borrows data from the non-concurrent periods to obtain the prior distribution for the control response in the concurrent periods, while accounting for the between-trial heterogeneity. Several frequentist and Bayesian methods for incorporating NCC data were recently compared in a comprehensive simulation study \cite{bofill2026treatment}.

Platform trials involve multiple hypothesis tests because several experimental treatments are investigated within a single trial. In addition, multiple endpoints or patient subgroups may be evaluated \cite{Collignon2020Master, Robertson2023Online}. In this setting, one may consider control of the platformwise type I error rate, defined as the probability of falsely rejecting at least one null hypothesis across the entire platform trial. This corresponds to the familywise error rate for the full set of hypotheses tested in the platform. However, it has been argued that no multiplicity adjustment is required for inferentially independent tests, such as for treatments with different mechanisms of action \cite{Koenig2024Current}. In that case, the pairwise error rate, that is, the probability of falsely rejecting the null hypothesis for a particular arm, needs to be controlled. Nevertheless, control of the familywise error rate at the substudy level remains necessary, when multiple doses, endpoints or subgroups are tested for a particular treatment.

Implementing a multiple testing procedure that controls the type I error rate at the platform level is challenging because, at the start of a platform trial, neither the total number of hypotheses nor the hypotheses themselves is typically pre-specified. Furthermore, results from the different arms become available only sequentially over time. Therefore, online multiple testing procedures have been proposed to control either the false discovery rate, defined as the expected proportion of false rejections among all rejections, or the family-wise error rate, defined as the probability of at least one false rejection within a family of hypotheses \cite{Zehetmayer2022Online, Robertson2023Online, greenstreet2024multi, greenstreet2025preplanned, fischer2025addis}. In such settings, false discovery rate control is of particular interest because it is less conservative than family-wise error rate control and may therefore provide greater power when many hypotheses are tested over time.
Interim analyses represent an additional source of multiplicity in group-sequential platform trials. Zehetmayer et al. \cite{Zehetmayer2022Online} proposed several methods for false discovery rate control in this setting and investigated their performance in a simulation study. Interim analyses in platform trials were also discussed by Greenstreet et al. \cite{greenstreet2024multi} in the context of a multi-stage design that allows late entry of new arms in a pre-planned manner. Assuming known entry times of the experimental arms and focusing mainly on settings without time trends, they proposed an approach for conducting interim analyses specified at the design stage while controlling the family-wise error rate. Related work by the same authors further considered pre-planned multi-stage platform trial designs in which multiple superior treatments may be identified \cite{greenstreet2025preplanned}. However, the approaches of Greenstreet et al. \cite{greenstreet2024multi, greenstreet2025preplanned} do not incorporate non-concurrent controls. Including interim analyses in platform trial designs that aim to use non-concurrent controls has not yet been discussed in the literature.

In group sequential trials, which allow for early termination of the trial due to efficacy or futility, it is known that the standard maximum likelihood estimates are generally no longer unbiased \cite{Wassmer2016Group, fan2004conditional}. This results from early stopping if more extreme results are observed in interim analyses. Many authors have proposed adjusted estimators to eliminate this bias \cite{grayling2023point, robertson2023point}. Aiming to reduce the mean bias, Whitehead \cite{whitehead1986bias} considered an adjustment strategy in which a new estimator is constructed from the original maximum likelihood estimator by subtracting the estimate of its bias. The resulting estimator is referred to as the mean adjusted estimator (MAE). This estimator offers a balanced trade-off in settings where equal importance is given to both bias and residual mean squared error, computed marginally as well as conditionally on stopping in a given stage \cite{grayling2023point}. Further improvement in the efficiency of unbiased estimators is often of interest. Emerson and Fleming proposed a derivation of a uniform minimum variance unbiased estimator (UMVUE) by applying the Rao-Blackwell theorem to an unbiased estimate of the treatment effect calculated from the data from the first stage \cite{emerson1990parameter}. The Rao-Blackwell theorem states that the variance of an unbiased estimator of an unknown parameter can be improved by conditioning on a sufficient test statistic for this parameter. Furthermore, according to the Lehmann-Scheffé theorem, if the considered test statistic is both sufficient and complete, the resulting Rao-Blackwellized estimator is a unique minimum variance unbiased estimator of the unknown parameter. Later, it was shown that the test statistic considered by Emerson and Fleming, assumed to be sufficient and complete, is indeed sufficient, but unfortunately not complete \cite{liu1999unbiased}. However, it was demonstrated that the proposed Rao-Blackwellized estimator is still UMVUE in a class of estimators that can be constructed in case of early stopping without requiring any knowledge of future analyses \cite{liu1999unbiased, grayling2023point, robertson2023point}. Another class of unbiased estimators is the so-called median unbiased estimators (MUE), which are constructed such that the probability of overestimating the true value of the unknown parameter is the same as the probability of underestimating it. In order to derive a median unbiased estimator in group sequential trials, a sample space ordering has to be chosen upfront, as the resulting MUE depends on this ordering. MUEs based on the stage-wise ordering \cite{emerson1990parameter}, likelihood ratio ordering \cite{chang1989confidence}, or the score test ordering \cite{rosner1988exact} can be constructed. A recent review by Grayling and Wason \cite{grayling2023point} provides a detailed overview of nine point estimators proposed for group sequential designs and compares them within a common framework for a two-stage group sequential trial. The authors argue that the optimal estimator depends on the operating characteristics that one desires to minimize in a given trial, and their importance in the marginal and conditional value.

Even though the bias in treatment effect estimates in classical group sequential trials has been well studied, the impact of conducting an interim analysis on the effect estimation in a platform trial utilizing non-concurrent controls in the analysis has not yet been investigated. In this work, we aim to examine how, in platform trials that evaluate multiple experimental treatments, the effect estimation in late-entering arms is affected by interim results in earlier arms.
For simplicity, we focus on a two-arm platform trial, where the second arm enters the trial later on, and an interim analysis for Arm 1 is performed at the time when Arm 2 is added. Such a timing of the interim analysis enables monitoring of the ongoing trial at the time of addition of the new arm \cite{greenstreet2024multi}. We examine the previously proposed model-based approach for including non-concurrent controls that includes time as a fixed effect \cite{Lee2020Including, Bofill2022Model}. Focusing on platform trials with continuous endpoints, we describe how the weight of non-concurrent controls included in the estimation of the treatment effect of Arm 2 depends on the interim result in Arm 1, and show that applying the current regression model, as considered by Lee and Wason \cite{Lee2020Including} and Bofill Roig et al. \cite{Bofill2022Model}, in group sequential platform trials leads to biased treatment effect estimators and type I error rate inflation. Following the idea from Whitehead \cite{whitehead1986bias}, we propose a mean adjusted estimator to mitigate this bias. 
We derive the analytical expression for the bias introduced due to the interim analysis enabling futility and efficacy stopping, and investigate the performance of the newly proposed estimator in terms of the bias, root mean squared error, type I error rate, and power in a simulation study.

The remainder of this paper is structured as follows: Section \ref{sec_designsetting} describes the considered design setting and reviews the current model-based approach that includes periods as a fixed effect. In Section \ref{sec_max_bias}, we explore the marginal and conditional bias that may arise when using the current regression model without any adjustments in a platform trial with an interim analysis. In Section \ref{sec_MAE}, we propose a treatment effect estimator that mitigates the bias due to interim analysis and consider the corresponding hypothesis test using this newly proposed estimator. In Section \ref{sec_simstudy}, we investigate the performance of the proposed estimators using simulations and compare them with the current methods. We conclude the paper with a discussion in Section \ref{sec_discussion}.

%%%%%%%%%%%%%%%%%%%%%%%%%%%%%%%%%%%%%%%%%%%%%%%%%%%%%%%%%%%%%%%%%%%%%%%%%%%%%%%%%%%%%%%%%%%%%%%

\section{Design Setting}\label{sec_designsetting}

\begin{figure}[h!]
    \centering
    \includegraphics[width=0.7\textwidth]{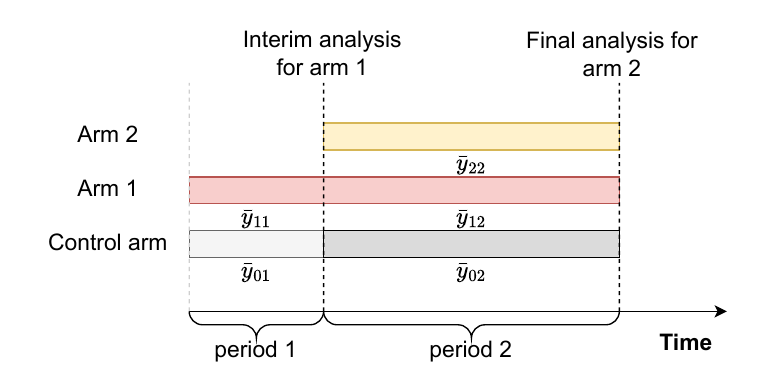}
    \caption{The considered platform trial design with an interim analysis for Arm 1 at the time point where Arm 2 joins the platform.}
    \label{fig:trial_design}   
\end{figure}

Consider a platform trial with $2$ treatment arms (indexed by $k=1,2$) and a common control group ($k=0$), where Arm 1 starts at the beginning of the trial, and Arm 2 enters the ongoing trial later on. Both arms finish the trial at the same time. The trial is split into two periods, divided by the time point where Arm 2 joins the platform. For Arm 1, an interim analysis is performed, with the possibility of stopping this arm due to futility or efficacy. It is assumed that this interim analysis is conducted at the time point of adding Arm 2. The second arm is only assessed in the final analysis at the end of the trial, after the total sample size is reached. In this final analysis, we aim to compare the efficacy of Arm 2 with the shared control group, i.e., including the NCC data for this arm. 
Direct pooling of NCC and CC data is statistically invalid in the presence of time trends; when unaccounted for, time trends can introduce severe bias into the treatment effect estimator \cite{Dodd2021Platform, Collignon2021Collaborative, Viele2014Use}. In this work, we therefore utilize a frequentist model-based approach to incorporate the NCC data into the analysis of Arm 2 while adjusting for time trends \cite{Bofill2022Model, Lee2020Including}.
The considered design is illustrated in Figure \ref{fig:trial_design}.

Before discussing trials involving an interim analysis, we review the model-based approach for analyzing platform trials using non-concurrent controls without interim analyses \cite{Bofill2022Model, Lee2020Including}, and describe the estimation of the treatment effect in Arm 2 compared to the shared control. Subsequently, we aim to adjust this estimate for an interim analysis conducted in Arm 1.

\subsection{Regression model with period adjustment in trials without interim analysis}\label{sec_reg_model_without_IA}

For the final analysis of Arm 2, we use a frequentist regression model that is fitted using all available data from the trial and adjusts for time trends by including the factor period as a categorical covariate \cite{Lee2020Including, Bofill2022Model}. This model is defined as follows:

\begin{equation*}\label{eq_freqmodel}
    E(y_j) = \eta_0  +  \sum_{k=1,2} \theta_k \cdot I(k_j = k)  +  \tau \cdot I(s_j=2)
\end{equation*}

where $y_j$ is the continuous response of patient $j$ ($j = 1, \ldots, N$), with $N$ denoting the total sample size in the trial. The intercept $\eta_0$ represents the control response in period 1, $\theta_k$ denotes the treatment effect of arm $k$ and $\tau$ indicates stepwise effect of period 2. The response variance, denoted by $\sigma^2$, is common for all arms and is assumed to be known.

Let $\bar y_{ks}$ denote the sample mean in arm $k$ and period $s$. The regression model estimates the effect of the treatment Arm 2 as the difference between the sample mean from this treatment arm and the model-based estimate of the control response in period 2. This estimator is then given by:

\begin{equation}\label{eq_trt2_estimate}
    \tilde\theta_2 = \bar{y}_{22} - \tilde y_{02}
\end{equation}

The model-based estimate of the control response in period 2 $\tilde y_{02}$ is a weighted average of the mean of the concurrent controls and the mean of the non-concurrent controls, adjusted by the time trend estimated from Arm 1:

\begin{equation*}\label{eq_ctrl_estimate}
    \tilde y_{02} = (1- \varrho) \cdot \bar{y}_{02} + \varrho \cdot [  \bar{y}_{01} + \bar y_{12} - \bar y_{11} ]
\end{equation*}

The weight given to the non-concurrent controls is given by the factor $\varrho$, which takes into account the sample sizes in each arm and period \cite{Bofill2022Model}:

\begin{equation}\label{eq_rho}
    \varrho = \frac{ \frac{ 1 }{ n_{02} } }{ \frac{ 1 }{ n_{01} } + \frac{ 1 }{ n_{02} } + \frac{ 1 }{ n_{11} } + \frac{ 1 }{ n_{12} } }
\end{equation}
where $n_{ks}$ denotes the sample size in arm $k$ and period $s$.

It was shown that in trials without interim analyses, \eqref{eq_trt2_estimate} is an unbiased treatment effect estimate if the time trends in all arms are equal and additive on the model scale \cite{Bofill2022Model}. 

\section{Bias introduced by an interim analysis in regression-based estimators}\label{sec_max_bias}

In this section, we investigate the bias of the model-based estimator \eqref{eq_trt2_estimate} if it is applied in a platform trial where an interim analysis is performed for Arm 1.

Suppose that an interim analysis of Arm 1 is performed at the time when Arm 2 is added to the trial. This interim analysis is based on a z-test comparing the mean responses between Arm 1 and control in period 1, i.e., testing $H_0: \theta_1=0$ vs $H_1: \theta_1 > 0$. We consider stopping for futility if the $p$-value from the z-test is larger than a futility bound $\alpha_F$ ($0 \le \alpha_F \le 1$), and stopping for efficacy if it is below an efficacy boundary $\alpha_E$ ($0 \le \alpha_E \le 1$, $\alpha_E < \alpha_F$). 
Note that when including an interim analysis for Arm 1, the treatment effect estimator for Arm 2 depends on the interim decision. If Arm 1 continues in the trial, the estimator given by \eqref{eq_trt2_estimate} depends on the non-concurrent control data, since by \eqref{eq_rho}  $\varrho>0$. However, if Arm 1 stops at the interim analysis, $n_{12}=0$ and therefore the weight $\varrho$ of the non-concurrent control data becomes zero and the model does not utilize the NCC data. In this case, the treatment effect estimate of Arm 2 is based only on the CC data: $\tilde{\theta}_2 = \bar y_{22} - \bar{y}_{02}$. 
Thus, whether non-concurrent control data are borrowed in the final estimator for Arm 2 is determined by the interim decision for Arm 1. Since this interim decision depends on the interim treatment effect estimate for Arm 1, it also depends on the non-concurrent control data. This dependence induces bias. Below, we derive the marginal and conditional bias of the model-based treatment effect estimator for Arm 2 for the platform trial design shown in Figure \ref{fig:trial_design}.

Assume patient $j$ enters the trial at time $t_j$, and is enrolled in period $s_j$. Further, assume a time trend defined by a function $f(t_j)$ of the patient entry time $t_j$ ($j=1, \ldots, N$), which is the same for all arms and additive on the model scale, as described in Bofill Roig et al. \cite{Bofill2022Model}.  For each period $s$, define $m_s := E[f(t_j) \mid s_j=s]$ and assume that $E[f(t_j) \mid k_j=k, s_j=s] = m_s$ for all arms $k$ that are open for randomization in period $s$. We additionally assume that the variances of the outcomes $y_j$ are equal across periods. This approximately holds when the strength of the time trend is small relative to the response variance conditional on time. Hence, the sample means in period 1 $\bar y_{01}$ and $\bar y_{11}$ follow normal distributions with means $\mu_0 + m_1$ and $\mu_1 + m_1$, respectively. In period 2, the sample means $\bar y_{02}$, $\bar y_{12}$, and $\bar y_{22}$ follow normal distributions with means $\mu_0 + m_2$, $\mu_1 + m_2$, and $\mu_2 + m_2$, respectively. Note that the treatment effects of arms 1 and 2, denoted by $\theta_1 = \mu_1 - \mu_0$ and $\theta_2 = \mu_2 - \mu_0$, respectively, are constant across the whole trial, since the time trend affects all arms equally. Denote by $Z_{11} = (\bar y_{11} - \bar y_{01}) / (\sigma \sqrt{\frac{1}{n_{11}} + \frac{1}{n_{01}}})$ the interim Z-statistic for Arm 1. The arm is stopped for futility if the interim p-value $p_{11} = 1- \Phi(Z_{11})$ is larger than a futility boundary $\alpha_F$, hence if $Z_{11} < \Phi^{-1}(1-\alpha_F) = c_F$. Similarly, the arm stops for efficacy if the p-value is below an efficacy boundary $\alpha_E$, i.e., $Z_{11} > \Phi^{-1}(1-\alpha_E) = c_E$, where $c_E > c_F$. Under the true treatment effect $\theta_1$, the probability to stop Arm 1 at the interim analysis is $\probP(Z_{11} < c_F) + \probP(Z_{11} > c_E) = \Phi(c_F - \delta) + \Phi(\delta - c_E)$, with $\delta =\theta_1/(\sigma\sqrt{\frac{1}{n_{11}} + \frac{1}{n_{01}}})$. Note that under the null hypothesis for Arm 1, this probability simplifies to $(1-\alpha_F) + \alpha_E$.

Since the inclusion of NCC data now stochastically depends on the interim outcome, $\tilde \theta_2$ is biased. As shown in Appendix \ref{app_bias}, its marginal bias --that is, the expected bias when averaging over interim outcomes (Arm 1 stopping in the interim analysis and continuing to period 2)-- is given by:

\begin{align}\label{eq_bias_marg}
    E[ \Tilde{\theta}_2  - \theta_2] = \varrho \cdot \sigma\sqrt{\frac{1}{n_{11}} + \frac{1}{n_{01}}} \cdot \left( \phi \left(c_F - \delta \right) - \phi \left( c_E - \delta \right) \right)
\end{align}
where $\phi(x)$ is the probability density function of the standard normal distribution evaluated at $x$.
Note that the bias does not depend on the time trend function $f(t_j)$, as long as the randomization is independent of the entry times $t_j$ in each period, since the bias is not caused by the time trend but arises from the interim analysis. 
Moreover, conditional on the event that Arm 1 stops, $\tilde{\theta}_2$ is unbiased. This follows since, in this case, the treatment effect estimate in Arm 2 does not depend on the data from period 1. Hence, the bias of $\tilde \theta_2$ conditional on the event that Arm 1 continues is given by the marginal bias \eqref{eq_bias_marg} divided by the probability that Arm 1 continues:

\begin{align}\label{eq_bias_cond}
   E[ \Tilde{\theta}_2  - \theta_2 | c_F \le Z_{11} \le c_E] = \frac{E[ \Tilde{\theta}_2  - \theta_2]}{\Phi(c_E - \delta) - \Phi(c_F - \delta)}
\end{align}
where $\Phi(x)$ is the cumulative distribution function of the standard normal distribution evaluated at $x$. The extent of this bias depends on: the weight of the NCC data when Arm 1 continues, $\varrho$ (i.e., the preplanned sample sizes in Arm 1 and control); the futility and efficacy boundaries, $\alpha_F$ and $\alpha_E$; and the standardized effect size in Arm 1, $\delta = \theta_1 / (\sigma\sqrt{\frac{1}{n_{11}} + \frac{1}{n_{01}}})$.
The model-based estimator is biased under both the null and alternative hypotheses for Arm 1. However, the bias does not depend on the treatment effect in Arm 2, $\theta_2$. Note that under the null hypothesis for Arm 1 ($\theta_1 = 0$), \eqref{eq_bias_cond} simplifies to:

\begin{align}\label{eq_bias_cond_null}
    E[ \Tilde{\theta}_2  - \theta_2 | c_F \le Z_{11} \le c_E] = \varrho \cdot \sigma\sqrt{\frac{1}{n_{11}} + \frac{1}{n_{01}}} \cdot \frac{\phi \left( c_F \right) - \phi \left( c_E \right)}{\alpha_F - \alpha_E}
\end{align}

\begin{figure}
    \centering
    \includegraphics[width=\linewidth]{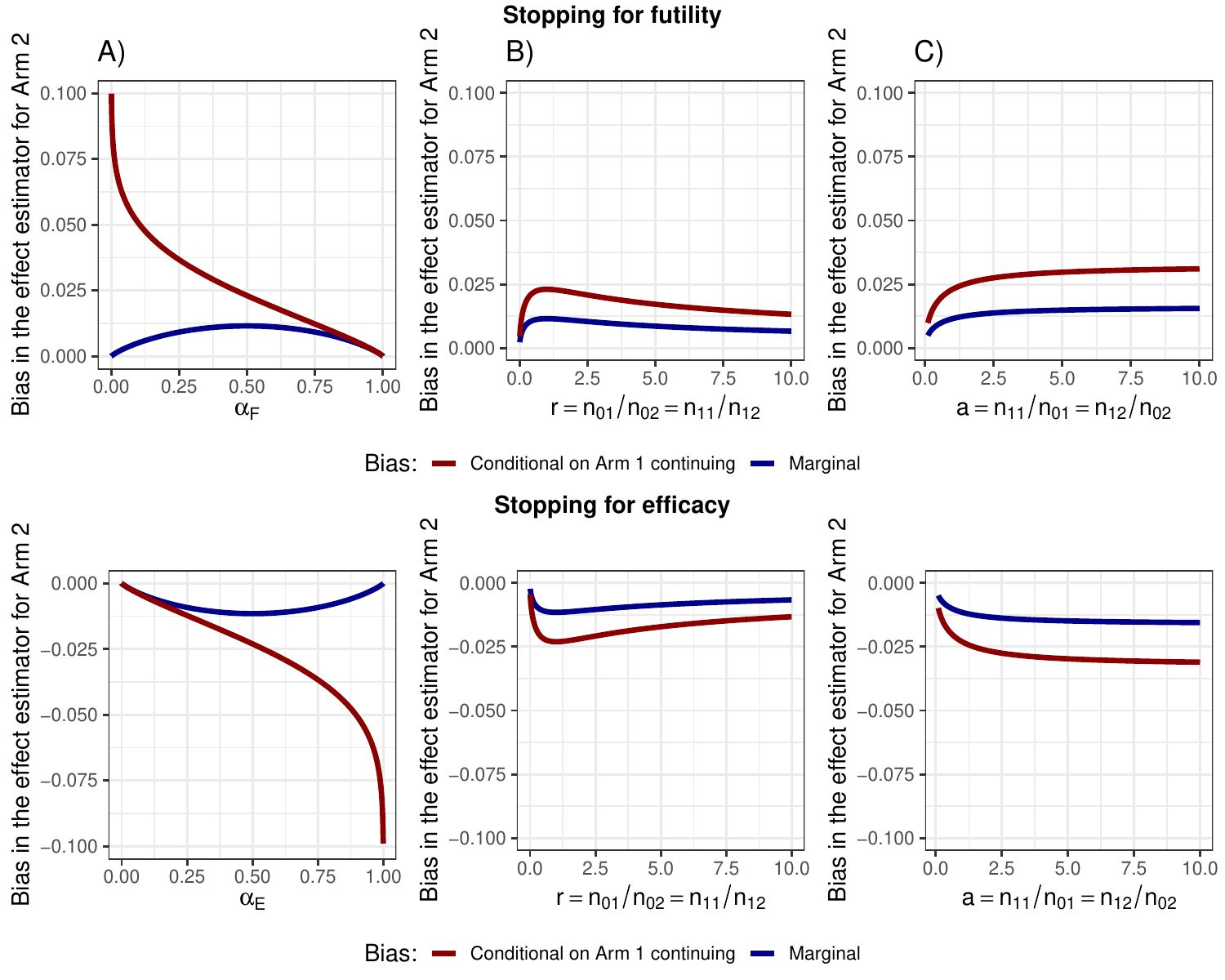}
    \caption{Marginal and conditional bias of the unadjusted model-based treatment effect estimator for Arm 2 \eqref{eq_trt2_estimate} when varying different design parameters. In all cases, no treatment effect for Arm 1 is assumed ($\theta_1=0$), and a unit variance $\sigma^2=1$ in each arm is used. In the first row, only stopping for futility is applied ($\alpha_E = 0$ in all cases), while in the second row, only efficacy stopping is used ($\alpha_F = 1$ in all cases). \\ \textbf{A)} Varying the futility bound $\alpha_F$ and $\alpha_E$. Sample sizes are set to 150 per arm and period. \\ \textbf{B)} Varying the sample sizes ratio between period 1 and period 2 $r = n_{01}/n_{02} = n_{11}/n_{12}$. Sample sizes per arm in period 2 are fixed ($n_{02} = n_{12} = 150$). Stopping bounds $\alpha_F = 0.5$ (first row) and $\alpha_E = 0.5$ (second row) are used. \\ \textbf{C)} Varying the allocation ratio in Arm 1 vs control $a = n_{11}/n_{01} = n_{12}/n_{02}$. The sample size in the control arm is fixed to 150 in each period ($n_{01} = n_{02} = 150$). Stopping bounds $\alpha_F = 0.5$ (first row) and $\alpha_E = 0.5$ (second row) are used.}
    \label{fig:maxbias}
\end{figure} 

Figure \ref{fig:maxbias} shows the marginal \eqref{eq_bias_marg} and conditional \eqref{eq_bias_cond} bias when varying different design parameters of the considered platform trial. If only futility stopping is applied (first row in Figure \ref{fig:maxbias}), the unadjusted model-based estimator is positively biased. This happens due to the underestimation of the mean response in the non-concurrent controls in cases where Arm 1 does not stop at the interim analysis, which results in the overestimation of the treatment effect in Arm 2. In contrast, with interim analyses allowing early stopping for efficacy only (second row in Figure \ref{fig:maxbias}), the unadjusted model-based estimator is, by symmetry, negatively biased. This occurs because the mean response in the non-concurrent controls is overestimated when Arm 1 continues, leading to underestimation of the treatment effect in Arm 2. 
Figure \ref{fig:maxbias}-A shows the bias as a function of the futility bound $\alpha_F$ (first row), or of the efficacy bound $\alpha_E$ (second row), when $\theta_1 = 0$. The absolute conditional bias increases with decreasing probability for Arm 1 to continue, i.e., with decreasing $\alpha_F$, and with increasing $\alpha_E$.
The absolute marginal bias is maximized for $\alpha_F = 0.5$ or $\alpha_E = 0.5$, i.e., when Arm 1 has a 50\% probability to continue after the interim analysis. Note that for $\alpha_F=0$ and $\alpha_F=1$, as well as $\alpha_E=1$ and $\alpha_E=0$, the treatment effect estimator $\tilde{\theta}_2$ is marginally unbiased, since these bounds result either in always using the separate analysis (since Arm 1 always stops) or always using the model-based approach (since Arm 1 always continues). The behavior of the bias with respect to the sample size ratio between period 1 and period 2 is shown in Figure \ref{fig:maxbias}-B. In particular, the sample sizes per arm in period 2 are set to 150, and the ratio $r = n_{01}/n_{02} = n_{11}/n_{12}$ is varied.  
Both the absolute marginal and conditional bias are maximized for $r=1$, i.e., when the periods are equally sized. This is because for smaller period 1 sample sizes ($r<1$), the weight of the period 1 data from Arm 1 in the Arm 2 treatment effect estimate is smaller. On the other hand, for larger period 1 sample sizes ($r>1$), the period 1 estimates have smaller standard errors and therefore their conditional bias (conditional on the event that the trial continues) is smaller. The impact of the allocation ratio between Arm 1 and control on the bias is illustrated in Figure \ref{fig:maxbias}-C. Specifically, the sample size in the control arm is fixed in each period ($n_{01} = n_{02} = 150$), and the sample size in Arm 1 is varied using the allocation ratio $a = n_{11}/n_{01} = n_{12}/n_{02}$. Note that the sample sizes per period are equal in each arm. Both marginal and conditional bias increase in their absolute value with the sample size in Arm 1 relative to the control sample size. 

\begin{figure}[h!]
    \centering
    \includegraphics[width=0.8\linewidth]{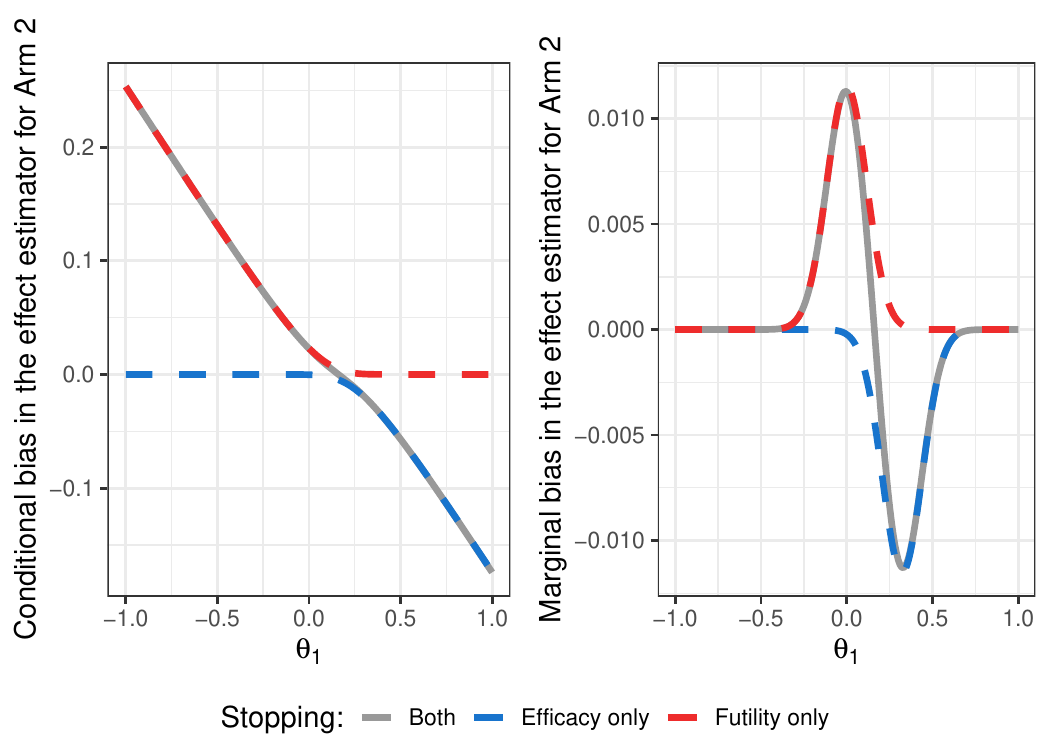}
    \caption{Conditional (left figure) and marginal (right figure) bias of the unadjusted model-based treatment effect estimator for Arm 2 \eqref{eq_trt2_estimate} when varying the treatment effect in Arm 1 ($\theta_1$). 
    In all cases, sample sizes of 150 in each arm and period are assumed, and unit variance $\sigma^2=1$ in each arm is used.
    Different lines show different stopping rules: futility stopping only ($\alpha_F=0.5$, $\alpha_E=0$, red line), efficacy stopping only ($\alpha_F=1$, $\alpha_E=0.00258$, blue line), and stopping for both futility and efficacy ($\alpha_F=0.5$, $\alpha_E=0.00264$, grey line). The chosen efficacy boundary $\alpha_E$ corresponds to the O'Brien-Fleming boundary using binding futility stopping at $\alpha_F$, assuming a significance level of 0.025 in the final analysis.
    Note that for better readability, the y-axis scale is different in each figure.}
    \label{fig:maxbias_theta}
\end{figure}

When both futility and efficacy stopping rules are applied, the marginal bias of the model-based estimator is influenced by both stopping rules, and its sign depends on the treatment effect in Arm 1 (see Figure \ref{fig:maxbias_theta}, grey line). For very small treatment effects in Arm 1, early stopping occurs predominantly for futility, resulting in a positive marginal bias in $\tilde{\theta}_2$. For very large treatment effects in Arm 1, early stopping occurs predominantly for efficacy, resulting in a negative marginal bias in $\tilde{\theta}_2$. For intermediate effect sizes, the opposing impacts of futility and efficacy stopping may partially offset one another, reducing the magnitude of the bias.

\clearpage

%%%%%%%%%%%%%%%%%%%%%%%%%%%%%%%%%%%%%%%%%%%%%%%%%%%%%%%%%%%%%%%%%%%%%%%%%%%%%%%%%%%%%%%%%%%%%%%

\section{An Estimator Adjusting for Bias Induced by the Interim Analysis}\label{sec_MAE}

In this section, we introduce a bias-adjusted treatment effect estimator for Arm 2 to mitigate the bias induced by the interim analysis in Arm 1.

\subsection{Mean adjusted estimator}

To account for the bias due to the interim analysis for Arm 1, we consider a mean adjusted estimator (MAE) \cite{whitehead1986bias}, constructed by subtracting the estimated conditional bias of $\Tilde{\theta}_2$ (conditional on the event that Arm 1 continues after the interim analysis) $E[ \Tilde{\theta}_2  - \theta_2 | c_F \le Z_{11} \le c_E]$ from the original model-based estimator, $\tilde \theta_2$, in cases when Arm 1 continues after the interim analysis.
The resulting bias-adjusted estimator of the treatment effect in Arm 2, $\tilde{\theta}^A_2$, is given by:
\begin{align}\label{eq_MAE}
    \tilde{\theta}^A_2 = \begin{cases} \bar{y}_{22} - \bar{y}_{02}, & \text{if } Z_{11} < c_F \text{ or } Z_{11} > c_E, \\[6pt] \tilde{\theta}_2 - \widehat{B}\left[\tilde{\theta}_2 \mid c_F \le Z_{11} \le c_E \right], & \text{if } c_F \le Z_{11} \le c_E, \end{cases}
\end{align}
with
\begin{align*}
\widehat{B}\left[\Tilde{\theta}_2  \mid c_F \le Z_{11} \le c_E \right]= \frac{\varrho \cdot \sigma\sqrt{\frac{1}{n_{11}} + \frac{1}{n_{01}}} \cdot \left(\phi \left(c_F - \hat{\delta} \right) - \phi \left(c_E - \hat{\delta} \right) \right)}{\Phi \left(c_E - \hat{\delta} \right) - \Phi \left(c_F - \hat{\delta} \right)} \text{, \hspace{0.2cm} where \hspace{0.2cm} } \hat{\delta} = \frac{\hat{\theta}_1}{\sigma \sqrt{\frac{1}{n_{11}} + \frac{1}{n_{01}}}}
\end{align*}
The term $\widehat{B}[\Tilde{\theta}_2 | c_F \le Z_{11} \le c_E ]$ denotes the estimator of the conditional bias defined in \eqref{eq_bias_cond}. 
Note that we use the conditional bias of $\tilde{\theta}_2$ for the adjustment, because the MAE is only employed in case Arm 1 continues, as there is no bias in the effect estimation for Arm 2 if Arm 1 stops at the interim.

The expression for the conditional bias \eqref{eq_bias_cond} depends on the futility bound and the standardized treatment effect of Arm 1. While the sample sizes, futility bound, and sample variance are considered as known parameters here, the treatment effect in Arm 1, $\theta_1$, is an unknown parameter. However, given that the mean adjusted estimator is computed at the end of the trial, we can estimate $\theta_1$ using the observed data. Therefore, in order to estimate $\widehat{B}[\Tilde{\theta}_2 | c_F \le Z_{11} \le c_E]$, we can use a plug-in estimator for $\theta_1$, denoted by $\hat \theta_1$. In particular, we consider four approaches for estimating $\theta_1$: (i) using the sample means of responses from Arm 1 and the control arm from the whole trial, denoted by $\bar y_{1\cdot}$ and $\bar y_{0\cdot}$ respectively, that is $\hat \theta_1 = \bar y_{1\cdot} - \bar y_{0\cdot}$; (ii) using only the sample means of responses from period 1, that is $\hat \theta_1 = \bar y_{11} - \bar y_{01}$; (iii) using only data from period 2, that is $\hat \theta_1 = \bar y_{12} - \bar y_{02}$; and (iv) the conditional uniform minimum variance unbiased estimator (CUMVUE), denoted by $\hat{\theta}_1^{CUMVUE}$, derived by applying the Rao-Blackwell theorem. The estimators (iii) and (iv) are unbiased estimators of $\theta_1$ conditionally on continuing after the interim.

As shown by Grayling and Wason \cite{grayling2023point}, the CUMVUE estimator for Arm 1 after period 2 can be constructed from the uniform minimum variance unbiased estimator (UMVUE), given by:

\begin{align}\label{eq_umvue}
\hat{\theta}_1^{UMVUE} = (\bar y_{1\cdot} - \bar y_{0\cdot}) - \frac{I_2 - I_1}{I_2 \sqrt{I_1}} \cdot \frac{\phi(c_E, Z_{12}\sqrt{I_1/I_2}, (I_2 - I_1)/I_2) - \phi(c_F, Z_{12}\sqrt{I_1/I_2}, (I_2 - I_1)/I_2)}{\Phi(c_E, Z_{12} \sqrt{I_1/I_2}, (I_2 - I_1)/I_2)-\Phi(c_F, Z_{12} \sqrt{I_1/I_2}, (I_2 - I_1)/I_2)}
\end{align}

where $Z_{12} = (\bar y_{1\cdot} - \bar y_{0\cdot}) / \sigma\sqrt{1/(n_{11} + n_{12}) + 1/(n_{01} + n_{02})}$ denotes the Z-statistic for arm 1 at the end of the trial, and $I_1 = 1/(\sigma^2(1/n_{11} + 1/n_{01}))$ and $I_2 = 1/(\sigma^2(1/(n_{11} + n_{12}) + 1/(n_{01} + n_{02})))$ are the information levels in periods 1 and 2, respectively. $\phi(x, \mu, \sigma^2)$ and $\Phi(x, \mu, \sigma^2)$ are the probability density and cumulative distribution functions of a normal distribution with mean $\mu$ and variance $\sigma^2$ evaluated at $x$, respectively. The CUMVUE for $\theta_1$ after period 2 is then constructed using \eqref{eq_umvue} as follows:

\begin{align*}
\hat{\theta}_1^{CUMVUE} = \frac{Z_{12} \cdot \sqrt{I_2} - I_1 \cdot \hat{\theta}_1^{UMVUE}}{ (I_2 - I_1)}
\end{align*}

\subsubsection{Hypothesis testing}

Now consider the problem of testing the null hypothesis of no treatment effect in Arm 2, $H_{0,2}: \theta_2 = 0$ against the one-sided alternative $H_{1,2}: \theta_2 > 0$. If Arm 1 continues after the interim analysis, we aim to utilize non-concurrent controls while adjusting for biases resulting from time trends and interim analysis in Arm 1. 
Therefore, we propose using the mean adjusted estimator, in \eqref{eq_MAE}, to define the Wald-type test statistic:

\begin{align*}\label{eq_test_stat}
    T = \begin{cases} \tilde{\theta}^A_2 / \sigma\sqrt{1/n_{02} + 1/n_{22}}, & \text{if } Z_{11} < c_F \text{ or } Z_{11} > c_E, \\[6pt] \tilde{\theta}^A_2 / \sqrt{\widehat{\text{Var}}(\tilde{\theta}^A_2 \mid c_F \le Z_{11} \le c_E}), & \text{if } c_F \le Z_{11} \le c_E, \end{cases}
\end{align*}
Under the null hypothesis $H_{0,2}$, the test statistic $T$ approximately follows the standard normal distribution. We reject $H_{0,2}$ at the one-sided significance level $\alpha$ if $T > \Phi^{-1}(1-\alpha)$. Note that for the case where Arm 1 stops at the interim analysis, this is equivalent to the standard z-test.

To obtain an estimator of the variance of the mean adjusted estimator $\tilde{\theta}^A_2$ in cases where Arm 1 continues, we employ the bootstrap. Specifically, the conditional variance $\text{Var}(\tilde{\theta}^A_2 \mid c_F \le Z_{11} \le c_E)$ is estimated via bootstrap stratified per arm and period. 
Letting $Y^{i*}_{ks}$ denote a response observation randomly drawn from arm $k$ and period $s$, we considered the following bootstrap algorithm:
\begin{enumerate}
\item Draw two independent bootstrap samples $Y^{1 *}_{01}, \ldots, Y^{n_{01} *}_{01}$ and $Y^{1 *}_{11}, \ldots, Y^{n_{11} *}_{11}$ from period 1 observations in control and Arm 1, respectively.

\item Compute the interim test statistic $Z_{11}^* = (\bar y^*_{11} - \bar y^*_{01}) / \sqrt{\frac{\sigma^2}{n_{11}} + \frac{\sigma^2}{n_{01}}}$ using data from the bootstrap sample. Here, $\bar y^*_{11}$ and $\bar y^*_{01}$ denote the mean of the bootstrap samples drawn in Step 1. 

\item If $Z_{11}^* < c_F$ or $Z_{11}^* > c_E$, discard this resample and restart at Step 1. Otherwise, continue to Step 4.

\item Draw three independent bootstrap samples $Y^{1 *}_{02}, \ldots, Y^{n_{02} *}_{02}$, $Y^{1 *}_{12}, \ldots, Y^{n_{12} *}_{12}$, and $Y^{1 *}_{22}, \ldots, Y^{n_{22} *}_{22}$ from period 2 of the trial; from control, Arm 1, and Arm 2, respectively.

\item Compute $\tilde{\theta}_{2}^{A*}$ with bootstrapped datasets from both periods.

\item Repeat Steps 1-5 until Arm 1 has continued after the interim analysis $B$ times, yielding estimates $\tilde{\theta}_{2,1}^{A*}, \ldots, \tilde{\theta}_{2, B}^{A*}$.

\item Compute the bootstrap variance of $\tilde{\theta}^A_2$ as follows:

$$\widehat{\text{Var}} \left( \tilde{\theta}^A_2 \mid c_F \le Z_{11} \le c_E \right) = \frac{1}{B} \sum_{j=1}^{B} \left( \tilde{\theta}_{2,j}^{A*} - \frac{1}{B} \sum_{j=1}^{B} \tilde{\theta}_{2,j}^{A*} \right)^2$$
\end{enumerate}

Note that the bootstrap procedure mimics the course of the original trial, preserving the sample sizes per arm and period, as well as the futility and efficacy stopping rules used at the interim analysis. In order to base the variance estimation on an equal number of bootstrap estimates (denoted by $B$), regardless of the futility and efficacy bounds $\alpha_F$ and $\alpha_E$, we repeat the bootstrap procedure until we obtain $B$ samples where Arm 1 has continued after the interim analysis.

%%%%%%%%%%%%%%%%%%%%%%%%%%%%%%%%%%%%%%%%%%%%%%%%%%%%%%%%%%%%%%%%%%%%%%%%%%%%%%%%%%%%%%%%%%%%%%%

\section{Simulation Study}\label{sec_simstudy}

\subsection{Evaluated operating characteristics}

We performed a simulation study to evaluate the properties of the proposed estimator and of the corresponding test statistic. 

In addition to examining the bias of the new estimator, it is also crucial to investigate its variance, quantified in terms of the root mean squared error, since unbiased or bias-reducing estimators oftentimes come with the price of increased variance. An estimator that, despite being unbiased, has a substantial increase in the root mean squared error compared to another approach should not be considered for practical use \cite{Wassmer2016Group}. Moreover, in group sequential designs, each operating characteristic can be considered in its marginal (overall) value, as well as conditional on stopping in a particular stage.

In the simulation study presented below, we therefore investigate the newly proposed estimator in terms of the bias and root mean squared error, and the corresponding hypothesis test in terms of type I error rate and statistical power. We discuss the marginal and conditional bias of both considered methods. For the remaining operating characteristics, we focus on their value conditional on Arm 1 continuing after the interim analysis, since the mean adjusted estimator is only employed in this case. Results for marginal operating characteristics can be found in the supplementary material (Section A). We compare the performance of the proposed mean adjusted estimate to the unadjusted model-based estimate and the separate analysis, which only considers data from period 2. All simulation scenarios were replicated 100.000 times to estimate the investigated operating characteristics. In the bootstrap algorithm for estimating the variance of $\tilde{\theta}^A_2$, we used $B=1000$ replicates.

\subsection{Data generation and considered design parameters}

We evaluate the performance of the proposed estimators in platform trials without time trends, as well as in trials that contain temporal drifts of different patterns and strengths.

Consider a platform trial with two experimental arms, as described in Section \ref{sec_designsetting} and illustrated in Figure \ref{fig:trial_design}. We assume uniform recruitment of one patient per time unit, such that $t_j=j$, and block randomization with block sizes of 4 and 6 in periods 1 and 2, respectively. The continuous outcome $y_j$ for patient $j$ is then drawn from a normal distribution according to $y_j \sim \mathcal{N}(\mu, \sigma^2)$ with 
\begin{gather*}
    \mu = \eta_0 + \sum_{k=1,2} \theta_k \cdot I(k_j = k) + f(j) \hspace{0.25cm} \text{and} \hspace{0.25cm} \sigma^2 = 1
\end{gather*}

where $\eta_0$ and $\theta_k$ are the response in the control arm and the effect of treatment $k$. 
The function $f(j)$ denotes the time trend with a magnitude parameterized by the parameter $\lambda$. In particular, we consider the following time trend patterns:

\begin{itemize}

    \item \textbf{Linear time trend:} $f(j) = \lambda \cdot \frac{j-1}{N-1}$, where $N$ indicates the maximum sample size in the trial, assuming that Arm 1 continues to period 2. The mean response in arm $k$ linearly increases with the slope $\lambda$ over time.
    
    \item \textbf{Stepwise time trend:} $f(j) = \lambda \cdot I(s_j = 2)$, where $s_j$ denotes the period for patient $j$. Hence, there is a jump in the mean response of size $\lambda$ when Arm 2 is added to the trial.

    \item \textbf{Seasonal time trend:} $f(j) = \lambda \cdot \mathrm{sin} \big( 2\pi \cdot \frac{j-1}{N-1} \big)$, where $N$ indicates the total sample size in the trial. The seasonal trend consists of one cycle, where the mean response increases at first and decreases afterwards, while the respective peaks of the cycle correspond to $\lambda$ or $-\lambda$.

    \item \textbf{Time trend following a random walk:} $\{ f(j) \}_{j=1}^N$, defined by $f(1) = 0$ and $f(j) = f(j-1) + \frac{\lambda}{N-1} \cdot \varepsilon_j$ for $j=2, \ldots, N$, where $\probP(\varepsilon_j = 1) = \probP(\varepsilon_j = -1) = 0.5$. Hence, the mean response for patient $j$ increases or decreases by $\frac{\lambda}{N-1}$ from the mean response of patient $j-1$, with probability 0.5, respectively. For patient $j=1$, no time trend is assumed.
\end{itemize}

Table \ref{tab_params} summarizes the values of the design parameters that we consider in the simulation study. 
In particular, the treatment effect in Arm 1 $\theta_1$ varies from -1 to 1.
The sample sizes per arm and period are varied in such a way that the impact of two ratios on the operating characteristics can be investigated. First, the ratio between the sample sizes in periods 1 and 2, $r = n_{01}/n_{02} = n_{11}/n_{12}$, which is varied from 1/15 to 10. Thus, the sample size per arm in period 1 can then be expressed as a multiple of the sample size per arm in period 2, which is kept constant. Second, the allocation ratio in Arm 1 and the control arm, denoted by $a = n_{11}/n_{01} = n_{12}/n_{02}$, is also varied from 1/15 to 10, allowing to examine the behavior of the methods under different allocation ratios. In particular, the sample size in Arm 1 is larger/smaller by a factor of $a$ with respect to the sample size in the control arm, which is kept constant. The sample sizes in Arm 2 and the control arm in period 2 are always set to 150. The strength of the time trend $\lambda$ is varied from $-0.15$ to $0.15$, i.e., considering both increase and decrease in the mean response over time.
In settings with varying $r$, $a$, and $\lambda$, the treatment effect $\theta_1$ was chosen such that a fixed design analysis of Arm 1 would result in approximately 80\% power.

\begin{table}[]
\centering
\begin{tabular}{|c|l|}
\hline
\multicolumn{1}{|l|}{\textbf{Design parameters:}}                                 & \textbf{Considered values:}                                                                                                                                       \\ \hline
Treatment effect $\theta_1$                                                       & \begin{tabular}[c]{@{}l@{}}[-1, 1]; with increments of 0.2 in [-1, -0.2] and [0.6, 1] \\ and increments of 0.05 in [-0.15, 0.5], \\ resulting in 22 values in total\end{tabular} \\ \hline
Ratio $r = n_{01}/n_{02} = n_{11}/n_{12}$                                         & 1/15, 1/3, \textbf{1}, 2, 4, 7, 10                                                                                                                                         \\ \hline
Ratio $a = n_{11}/n_{01} = n_{12}/n_{02}$                                         & 1/15, 1/3, \textbf{1}, 2, 4, 7, 10                                                                                                                                         \\ \hline
Time trend strength $\lambda$                                                     & -0.15, -0.075, \textbf{0}, 0.075, 0.15                                                                                                                                     \\ \hline
\multicolumn{1}{|l|}{\textbf{Sample sizes considered fixed:}}                     &                                                                                                                                                                   \\ \hline
$n_{02}$                                                                          & \textbf{150}                                                                                                                                                               \\ \hline
$n_{22}$                                                                          & \textbf{150}                                                                                                                                                               \\ \hline
\multicolumn{1}{|l|}{\textbf{Sample sizes resulting from the considered ratios:}} &                                                                                                                                                                   \\ \hline
$n_{01}$                                                                          & 10, 50, \textbf{150}, 300, 600, 1050, 1500                                                                                                                                 \\ \hline
$n_{11}$                                                                          & 10, 50, \textbf{150}, 300, 600, 1050, 1500                                                                                                                                 \\ \hline
$n_{12}$                                                                          & 10, 50, \textbf{150}, 300, 600, 1050, 1500                                                                                                                                 \\ \hline
\end{tabular}
\caption{Considered values of the design parameters varied in the simulations and the resulting sample sizes. When varying one design parameter, the remaining ones were kept constant at the value highlighted in bold type. In each case, the futility boundary was set to $\alpha_F = 0.5$, while the efficacy boundary $\alpha_E$ was set to the O'Brien-Fleming boundary for the given sample sizes and $\alpha_F$. In settings with varying $r$, $a$, and $\lambda$, the treatment effect $\theta_1$ was chosen such that a fixed design analysis of Arm 1 would have 80\% power.  In total, 41 scenarios were considered under both the null and alternative hypotheses. Note that not all combinations are considered.}
\label{tab_params}
\end{table}

We focus on testing the null hypothesis $H_{0,2}: \theta_2 = \mu_2 - \mu_0 = 0$ against the one-sided alternative $H_{1,2}: \theta_2 > 0$ at the significance level $\alpha=0.025$ using the proposed mean adjusted estimator $\tilde{\theta}^A_2$, as well as the unadjusted model-based estimator $\tilde{\theta}_2$. For comparison, we include results for the root mean squared error and statistical power obtained when using the separate analysis, regardless of the interim outcome. We consider scenarios where Arm 2 is under the null hypothesis ($\theta_2 = 0$), as well as cases where the alternative holds. In the latter case, the treatment effect for Arm 2 is set to $\theta_2 = 0.32$ such that the separate analysis comparing Arm 2 to the concurrent controls leads to approximately 80\% power using the given sample sizes.

\subsection{Results}

\begin{figure}[h!]
    \centering
    \includegraphics[width=\linewidth]{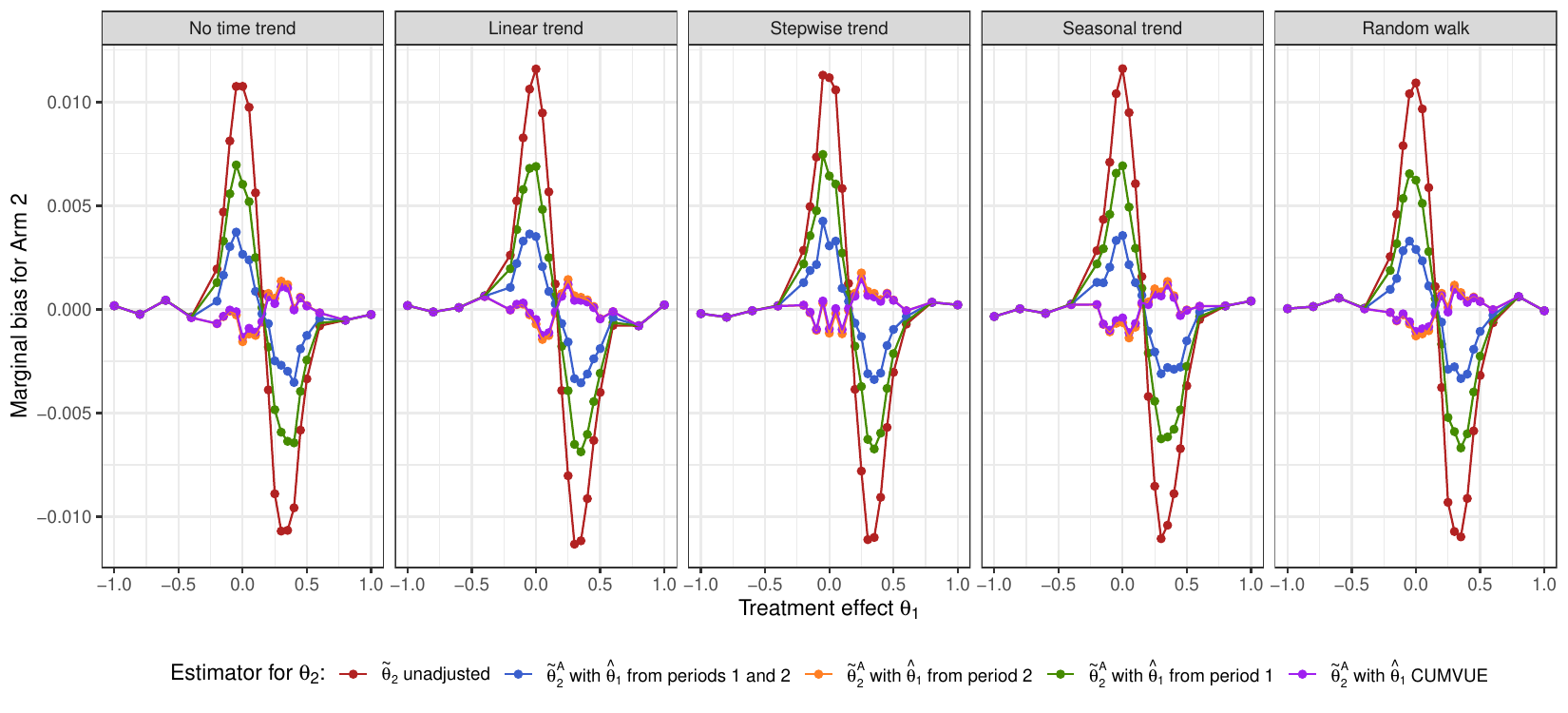}
    \caption{Marginal bias in the treatment effect estimator for Arm 2 for varying the treatment effect in Arm 1 ($\theta_1$) using the unadjusted estimator and the mean adjusted estimator with $\theta_1$ estimated from both periods, only period 1, only period 2, or using the CUMVUE. This figure corresponds to a platform trial with sample sizes per arm and period set to 150. The futility boundary was set to $\alpha_F = 0.5$, while for the efficacy boundary, $\alpha_E = 0.00264$ was chosen, corresponding to the O'Brien-Fleming boundary for the given sample sizes and $\alpha_F$, assuming a significance level of 0.025 in the final analysis. The strength of the time trend is $\lambda=0.15$ for all considered patterns.}
    \label{fig:marginal_bias_theta1}
\end{figure}

\begin{figure}[h!]
    \centering
    \includegraphics[width=\linewidth]{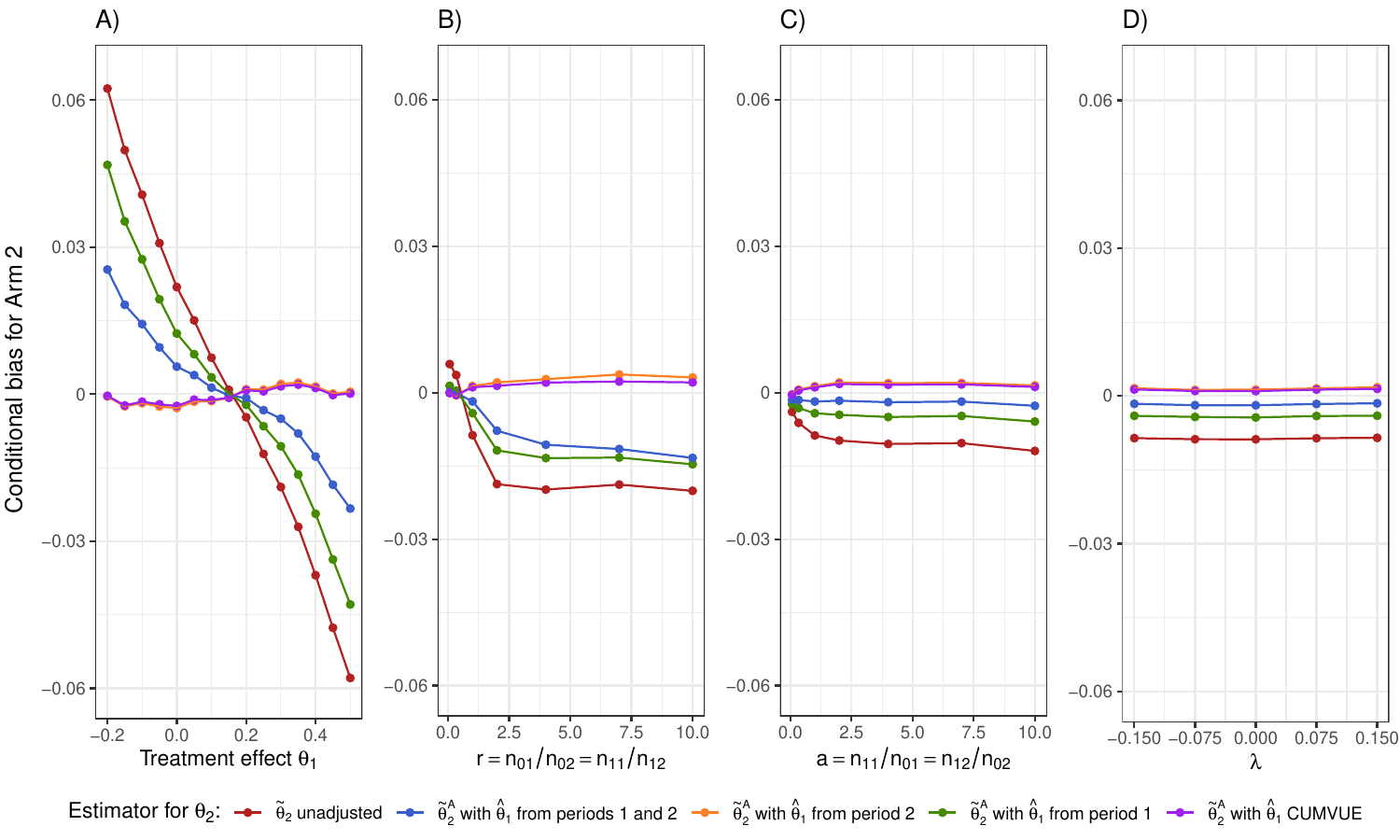}
    \caption{Conditional bias in the treatment effect estimator for Arm 2 using the unadjusted estimator $\tilde{\theta}_2$ and the mean adjusted estimator $\tilde{\theta}^A_2$ with $\theta_1$ estimated from both periods, only period 1, only period 2, or using the CUMVUE. \\ \textbf{A)} Varying treatment effect in Arm 1 ($\theta_1$). This figure corresponds to a platform trial without time trends, with sample sizes per arm and period set to 150. The futility boundary was set to $\alpha_F = 0.5$, while for the efficacy boundary, $\alpha_E = 0.00264$ was chosen, corresponding to the O'Brien-Fleming boundary for the given sample sizes and $\alpha_F$, assuming a significance level of 0.025 in the final analysis. Note that we have restricted the range of the values for $\theta_1$, such that Arm 1 has a reasonable probability to continue after the interim analysis ($>4\%$). \\ \textbf{B)} Varying ratio between the sample sizes in period 1 and period 2 ($r$), where the period 2 sample sizes per arm are fixed. Sample sizes per arm in period 1 are determined by the chosen ratio $r$, i.e., increase or decrease with respect to the period 2 sample sizes by a factor of $r$. This figure corresponds to a platform trial without time trends, with sample sizes per arm in period 2 set to 150. The futility boundary was set to $\alpha_F = 0.5$, while the efficacy boundary $\alpha_E$ was set in each case to the corresponding O'Brien-Fleming boundary for the given sample sizes and $\alpha_F$, assuming a significance level of 0.025 in the final analysis. The treatment effect in Arm 1 was in each case chosen such that a fixed design analysis of Arm 1 would yield approx. 80\% power. \\ \textbf{C)} Varying ratio between the sample sizes in Arm 1 and control ($a$), where the control arm sample sizes per period are fixed. Sample sizes in Arm 1 per period are determined by the chosen ratio $a$, i.e., increase or decrease with respect to the control arm sample sizes by a factor of $a$. This figure corresponds to a platform trial without time trends, with sample sizes in the control arm per period set to 150. The futility boundary was set to $\alpha_F = 0.5$, while for the efficacy boundary, $\alpha_E = 0.00264$ was chosen, corresponding to the O'Brien-Fleming boundary for the given sample sizes and $\alpha_F$, assuming a significance level of 0.025 in the final analysis. The treatment effect in Arm 1 was in each case chosen such that a fixed design analysis of Arm 1 would yield approx. 80\% power. \\ \textbf{D)} Varying strength of the time trend ($\lambda$). This figure corresponds to a platform trial with a linear time trend of strength $\lambda$, with sample sizes per arm and period set to 150. The futility boundary was set to $\alpha_F = 0.5$, while for the efficacy boundary, $\alpha_E = 0.00264$ was chosen, corresponding to the O'Brien-Fleming boundary for the given sample sizes and $\alpha_F$, assuming a significance level of 0.025 in the final analysis. The treatment effect in Arm 1 was in each case chosen such that a fixed design analysis of Arm 1 would yield approx. 80\% power.}
    \label{fig:cond_bias}
\end{figure}

We first examine the marginal and conditional bias, summarized in Figures \ref{fig:marginal_bias_theta1} and \ref{fig:cond_bias}. Figure \ref{fig:marginal_bias_theta1} shows the marginal bias in the effect estimator for Arm 2 for varying treatment effects in Arm 1, $\theta_1$, and under different time trend patterns. The bias is given for the unadjusted model-based estimator, as well as the proposed mean adjusted estimator with $\theta_1$ estimated from either both, only period 1, only period 2, or using the Rao-Blackwellized CUMVUE. As discussed in Section \ref{sec_max_bias}, the marginal bias of the unadjusted estimator $\tilde{\theta}_2$ reaches its maximum in the absolute value for $\theta_1 = 0$ and $\theta_1 = 0.3,$ i.e., in cases where Arm 1 continues in the trial with approx. 50\% probability. The mean adjusted estimators provide a substantial reduction in the marginal bias. In particular, this reduction is most pronounced for the MAEs with $\theta_1$ estimated only from period 2 or using the CUMVUE, even though they still maintain a slight bias in the extreme cases where Arm 1 continues with approx. 50\% probability. The MAEs with $\hat{\theta}_1$ estimated from both periods or only period 1 still maintain considerable bias due to the inclusion of the period 1 data, which was used for the interim decision. This is more pronounced when the period 1 data are used for the estimation of $\theta_1$ exclusively. The obtained bias is consistent across scenarios with no time trend, and linear, stepwise, seasonal, or random walk time trends with $\lambda=0.15$. Hence, the mean adjusted estimator is robust to time trends regardless of their pattern and strength. This holds for all considered operating characteristics. Therefore, in the main paper, we only present results under no time trends, and results for linear time trends. The corresponding results for other time trends are presented in the supplementary material (Section A).

Figure \ref{fig:cond_bias} shows the bias conditional on Arm 1 continuing after the interim analysis, as a function of the treatment effect in Arm 1 $\theta_1$, the ratio between sample sizes in periods 1 and 2 $r$, the ratio between sample sizes in Arm 1 and control $a$, and the strength of the time trend $\lambda$. Again, the conditional bias of the model-based estimator is reduced in all cases when using the mean adjusted estimator. This improvement is more pronounced in designs that  have more extreme treatment effects in Arm 1, since the conditional bias of the unadjusted estimator has the highest value in these cases (see Figure \ref{fig:cond_bias}-A). The MAEs with $\theta_1$ estimated from either both periods or only period 1 still show some conditional bias, which also decreases as $\theta_1$ becomes more moderate. 
This overestimation or underestimation of $\theta_1$ in period 1 leads to lower or higher values of the estimated bias used for the computation of the MAEs, if the period 1 data is included. As a result, there is insufficient bias reduction and hence a remaining positive or negative bias in the treatment effect estimation for Arm 2. 
The MAEs with $\theta_1$ estimated only from period 2 and using the CUMVUE are only very slightly negatively or positively biased. However, their performance is more robust to the varying $\theta_1$.  
Moreover, the conditional bias when using the MAEs with $\theta_1$ estimated from period 2 or employing the CUMVUE, does not depend sensitively on the sample size ratio between periods 1 and 2 (see Figure \ref{fig:cond_bias}-B), or the allocation ratio between Arm 1 and the control arm (see Figure \ref{fig:cond_bias}-C), and does not depend on the strength of the time trend $\lambda$ (see Figure \ref{fig:cond_bias}-D).

\begin{figure}[h!]
    \centering
    \includegraphics[width=\linewidth]{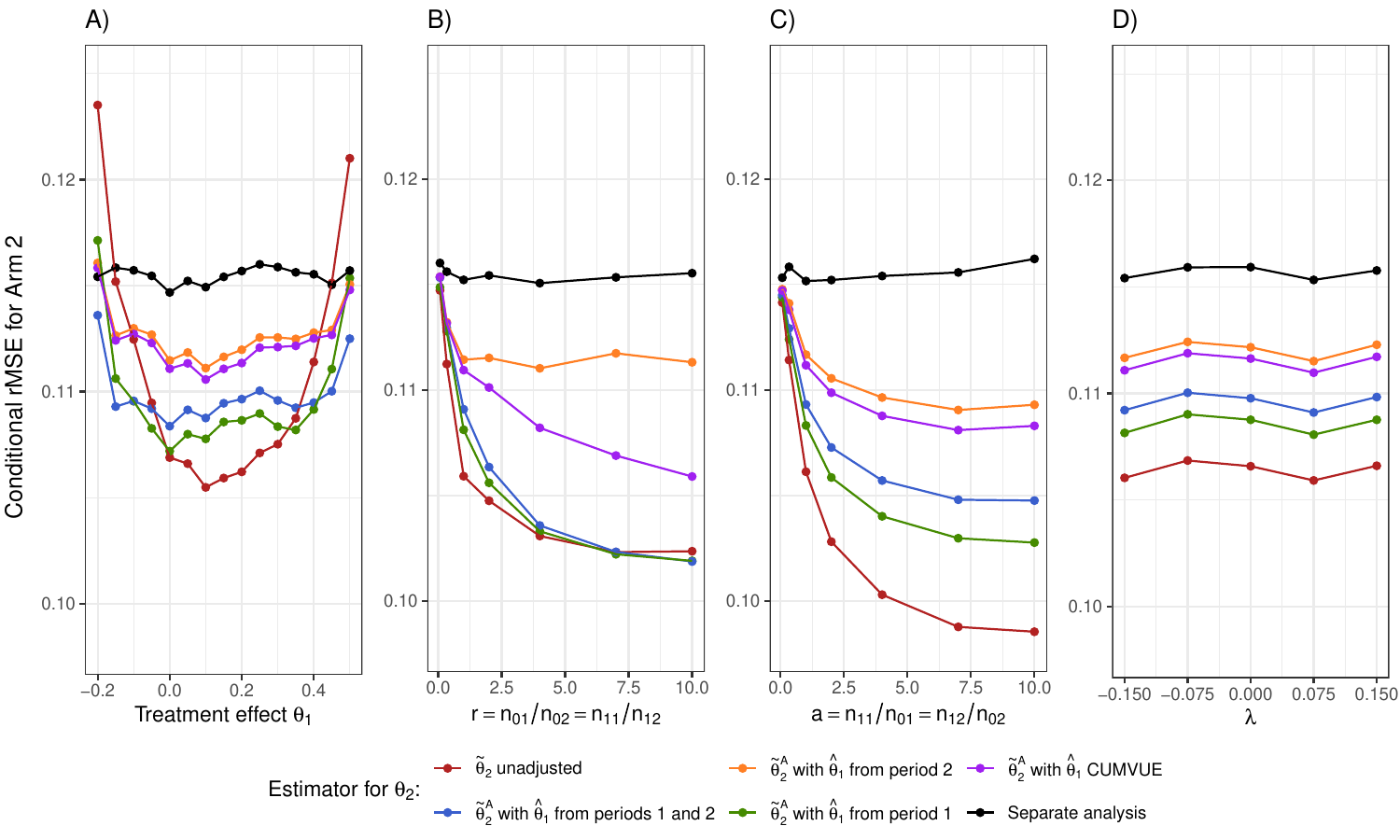}
    \caption{Conditional root mean squared error of the treatment effect estimator for Arm 2 using the unadjusted estimator $\tilde{\theta}_2$, the mean adjusted estimator $\tilde{\theta}^A_2$ with $\theta_1$ estimated from both periods, only period 1, only period 2, or using the CUMVUE, as well as using the separate analysis. See the legend of Figure \ref{fig:cond_bias} for the explanation of the subfigures \textbf{A)}-\textbf{D)}.} 
    \label{fig:cond_MSE}
\end{figure}

Figure \ref{fig:cond_MSE} shows the conditional root mean squared error (rMSE) of the newly proposed estimator compared to the model-based unadjusted estimator. In almost all scenarios, the adjusted estimators achieve a lower rMSE than the separate analysis using concurrent controls only. However, in most scenarios, the bias adjusted estimators based on the period 2 estimate of $\theta_1$ and the CUMVUE have a larger mean squared error than the unadjusted estimator. Exceptions are extreme cases, where Arm 1 has a very low probability to continue (see Figure \ref{fig:cond_MSE}-A), where the rMSE of the unadjusted estimator exceeds the rMSE of the mean adjusted estimators due to its very large bias in such settings. For all considered methods that take into account the data from period 1, the rMSE is decreasing with increasing sample sizes in period 1 (see Figure \ref{fig:cond_MSE}-B).

\begin{figure}[h!]
    \centering
    \includegraphics[width=\linewidth]{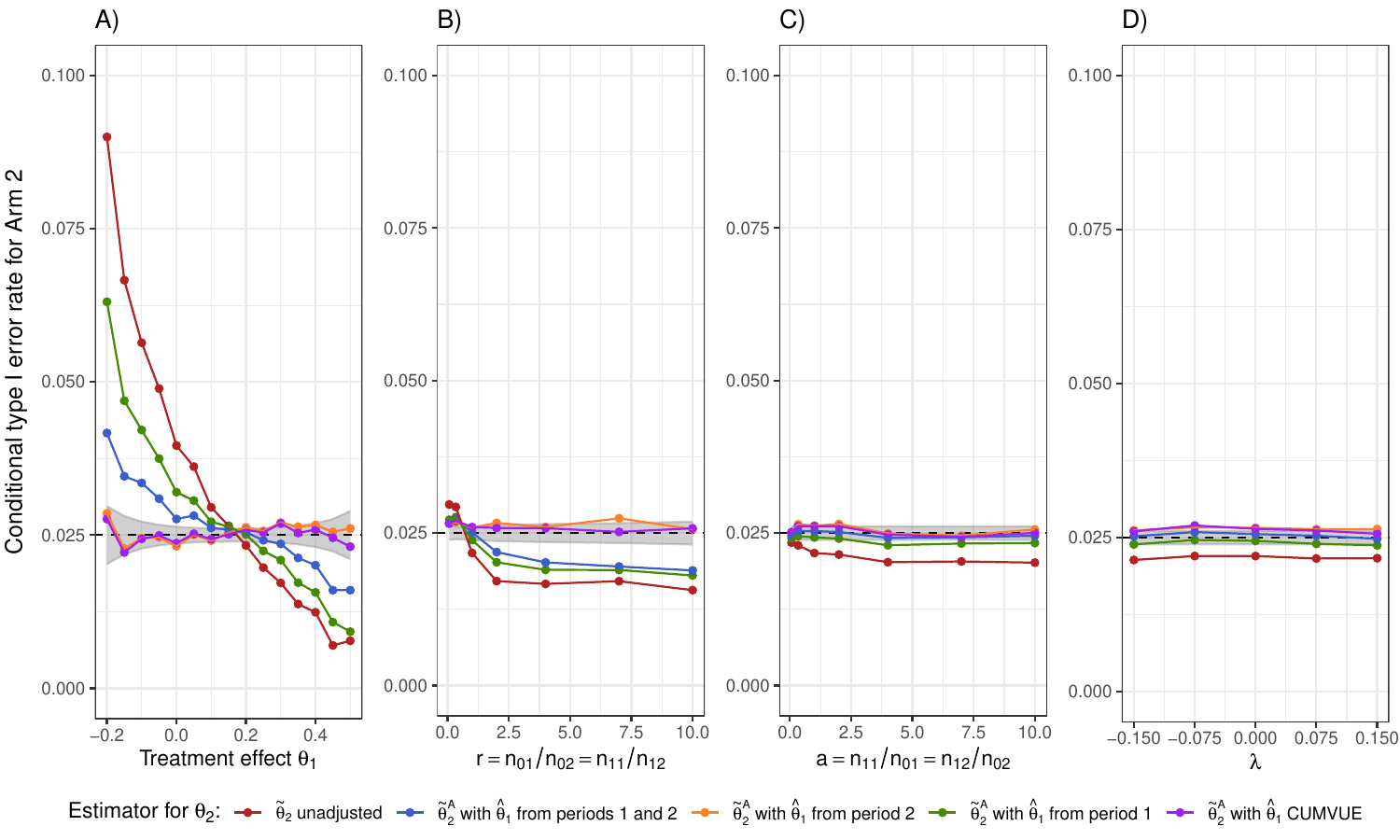}
    \caption{Conditional type I error rate for Arm 2 using the unadjusted estimator $\tilde{\theta}_2$ and the mean adjusted estimator $\tilde{\theta}^A_2$ with $\theta_1$ estimated from both periods, only period 1, only period 2, or using the CUMVUE. All figures include a dashed reference line for the nominal significance level of 0.025 and the simulation error is shown as a gray area representing the 95\% confidence interval of the simulated type I error rate. Note that in Figure \ref{fig:cond_t1e}-A, this area is not constant over different values of $\theta_1$. This is because for more extreme values of $\theta_1$, the conditional type I error rate is estimated based on fewer samples, since Arm 1 has a lower probability to continue after the interim analysis in these cases. See the legend of Figure \ref{fig:cond_bias} for the explanation of the subfigures \textbf{A)}-\textbf{D)}.} 
    \label{fig:cond_t1e}
\end{figure}

Next, we assess the properties of the corresponding hypothesis test. The conditional type I error rate is shown in Figure \ref{fig:cond_t1e}. The inflation or deflation of the conditional type I error rate caused by applying the unadjusted model-based estimator is substantially reduced when using the mean adjusted estimators. However, the MAEs computed using $\hat{\theta}_1$ from period 1 or both periods still show some inflation or deflation. The MAEs with $\theta_1$ estimated using the CUMVUE yield type I error rates within the simulation error in all considered scenarios. Figure \ref{fig:cond_t1e}-D demonstrates that the type I error rate for given sample sizes and stopping bounds is not dependent on the strength of the time trend.

\begin{figure}[h!]
    \centering
    \includegraphics[width=\linewidth]{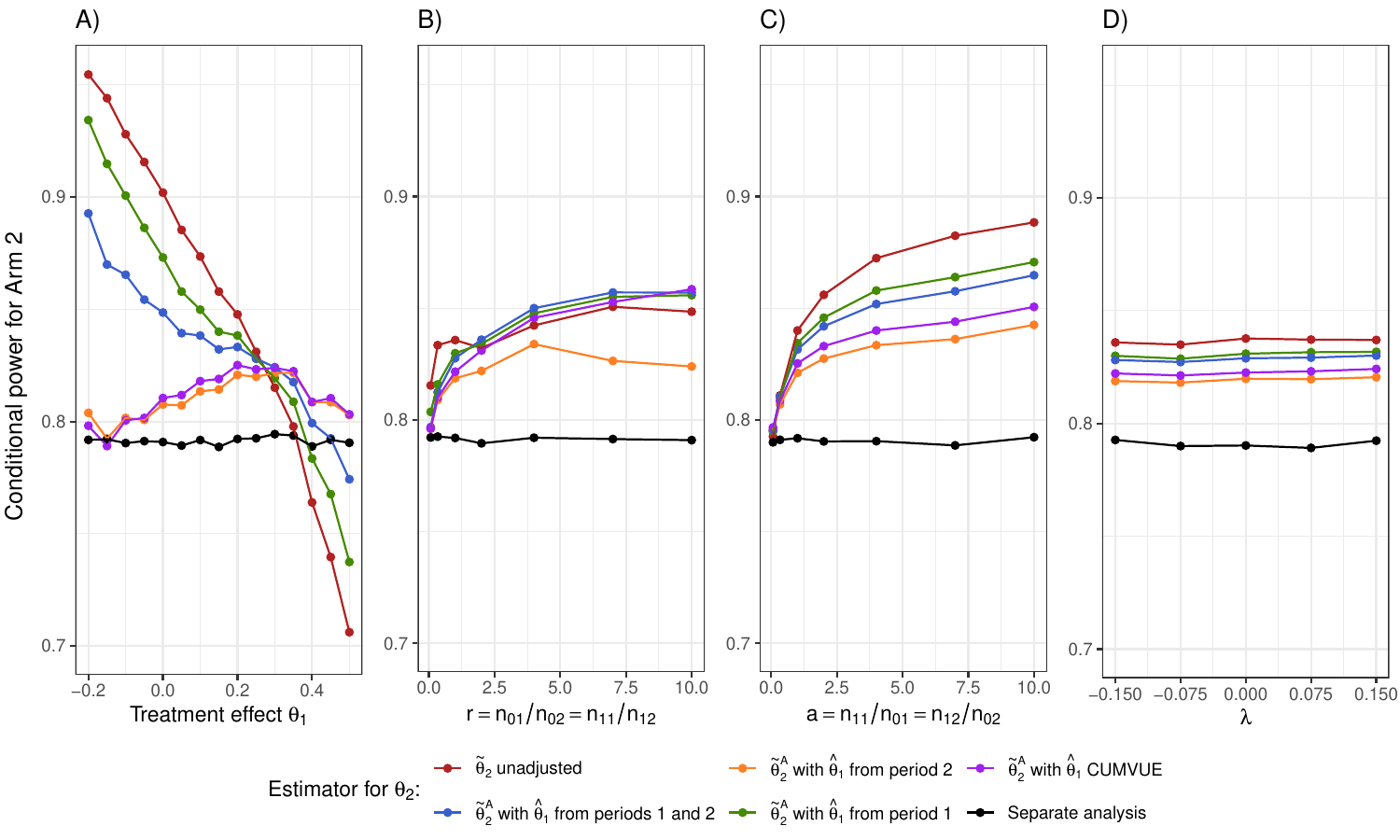}
    \caption{Conditional power for Arm 2 using the unadjusted estimator $\tilde{\theta}_2$, the mean adjusted estimator $\tilde{\theta}^A_2$ with $\theta_1$ estimated from both periods, only period 1, only period 2, or using the CUMVUE, as well as using the separate analysis. See the legend of Figure \ref{fig:cond_bias} for the explanation of the subfigures \textbf{A)}-\textbf{D)}.} 
    \label{fig:cond_pow}
\end{figure}

Figure \ref{fig:cond_pow} shows the conditional power of the model-based procedures and, for comparison, the separate analysis using only concurrent controls. The highest power is achieved by the unadjusted model-based estimator in most of the considered cases, which, however, is due to the bias in the treatment effect estimation and inflation in the type I error rate. Nevertheless, the mean adjusted estimators that control the type I error rate (based on $\theta_1$ estimates from period 2 and using the CUMVUE) also lead to power gains compared to the separate analysis, as they make use of the non-concurrent control data. As expected, the conditional power increases for increasing period 1 sample sizes (keeping the period 2 sample size fixed) and increasing sample sizes in Arm 1 (see Figures \ref{fig:cond_pow}-B and \ref{fig:cond_pow}-C, respectively). As the other operating characteristics, the conditional power does not depend on the strength of the time trend (see Fig. \ref{fig:cond_pow}-D). 

In the settings considered in this Section, early stopping occurs predominantly for efficacy, which leads to conservative treatment effect estimates and hypothesis tests. In the supplementary material (Section B), we present additional results for settings where only stopping for futility is applied. Such scenarios yield positive bias and type I error rate inflation in the unadjusted estimator, which is considerably reduced by applying the mean adjusted estimators. In particular, the MAE computed using the CUMVUE estimator for $\theta_1$ shows the best performance across the considered scenarios.

\clearpage

\section{Discussion}\label{sec_discussion}

Existing methods for NCC inclusion have not accounted for interim analyses. 
We demonstrated that model-based approaches used to adjust for time trends in platform trials can introduce a positive bias and inflate type I error rates once futility interim analyses are included, even in the absence of actual time trends. Similarly, efficacy stopping may lead to negative bias and deflation of the type I error rate. In this work, we showed that conducting an interim analysis in one arm affects the treatment effect estimation in other arms when non-concurrent controls are used. This occurs because the weight of the non-concurrent controls in the model-based estimator depends on the interim result of other arms.

From the analytical expression for the bias, we can see that inflated or deflated marginal type I error rates of the unadjusted model-based estimator for Arm 2 are most pronounced when (i) borrowing from NCC is substantial (i.e., large $\varrho$), (ii) the interim decision for Arm 1 has a moderate probability of stopping or continuing (i.e., continuation probability of around 50\%), and (iii) period 1 and period 2 sample sizes are similar. Conversely, the inflation/deflation is negligible when NCC borrowing is minimal (i.e., $\varrho \approx 0$, for example when there is a very large sample size for the concurrent control in period 2), when the interim rule is extreme (meaning that Arm 1 almost always stops or continues), or when periods 1 and 2 substantially differ in sample sizes.

To address this challenge in a platform trial with two experimental arms and two periods, we proposed mean adjusted treatment effect estimators for Arm 2, derived from the original model-based estimator by subtracting its estimated bias. This estimator adjusts for time trends while accounting for the interim analysis in Arm 1. 
Because the bias-adjusted estimator for Arm 2 depends on an estimate of the treatment effect $\theta_1$ of Arm 1, we considered several estimators. In a simulation study, we showed that using the CUMVUE to estimate $\theta_1$ minimized the bias of the bias-adjusted estimator of Arm 2 across all considered options.

Moreover, we proposed an adjusted hypothesis test for Arm 2, using the Wald test based on the mean adjusted estimator. This test controls the type I error rate in the considered scenarios when using the CUMVUE to estimate $\theta_1$ and yields power gains compared to a separate analysis using only concurrent controls, due to the increased sample size from incorporating non-concurrent controls. However, these power gains are notably smaller than those achieved by the regression model in settings without interim analyses. This is because the mean adjusted estimator accounts simultaneously for time trends and the interim analysis. When using non-concurrent controls, the potential gains in power must be weighed against the added complexity and reduced robustness to model assumptions.
The proposed estimator is robust to time trends regardless of their pattern. However, since the time trend adjustment relies on the assumptions of equal time trends across all arms and additivity of the time trends on the model scale, deviations from these assumptions may lead to bias and inflation of the type I error rate \cite{Bofill2022Model}.

If, in addition to the interim analysis for Arm 1, an interim analysis is also performed for Arm 2, this introduces an additional bias in the treatment effect estimate for Arm 2. An efficacy interim analysis for Arm 2 will also impact the type I error rate of hypothesis tests, requiring the use of adjusted boundaries. 

We considered a two-arm platform trial to illustrate the bias introduced by including non-concurrent control data, which is caused by the dependence of the weight of the NCC data in the treatment effect estimator on the interim results of previous arms. In our example, the interim analysis for Arm 1 is performed at the time Arm 2 is added. This timing of the interim analysis is the most extreme scenario, as the impact of the interim analysis on the weight of the non-concurrent control data is maximized: if Arm 1 stops early, the NCC data are not used at all in the estimate. If the interim analysis were performed at a later time point, the weight of the non-concurrent controls would remain strictly positive in case of stopping when Arm 1 stops.

Recent methodological work has identified a related issue in platform trials that arises when allowing for a change in the control arm. For instance, when an experimental treatment is found to be superior and becomes the new standard of care. In this setting, a decision must be made whether to keep or discard the information collected before the control is changed. Greenstreet et al. show that this choice can also substantially alter the conditional type I error rate and conditional power \cite{greenstreet2025design}.

In this work, we assumed that the final analysis of both experimental arms is performed at the same time. The resulting 2-period design represents an extreme case with respect to the bias of the treatment effect estimator for Arm 2, because it maximizes the overlap between the two experimental arms. A larger overlap increases the contribution of non-concurrent controls in the estimator from the period-adjusted regression model, whereas no overlap implies no borrowing. As a result, the interim decision on Arm 1 has a stronger impact on the Arm 2 estimator in this setting. If Arm 1 has an earlier final analysis, the trial includes an additional third period with only Arm 2 and control, thereby reducing the overlap between the two experimental arms.

Extending the proposed methodology to incorporate more general timings of interim and final analyses, trials with more than two experimental arms, and interim analyses for all arms is a topic for future research. These extensions would require accounting for additional complexity in how NCC data contributes to the estimator for late-entering arms. In particular, if interim or final analyses are conducted when no additional arms are entering the trial, the trial is split into more periods. Each additional period introduces a new period effect in the regression model, and the weight assigned to the NCC data is determined by the amount of overlapping sample size between experimental arms. Moreover, in platform trials with more than two experimental arms, the treatment effect estimator for late-entering arms may depend on interim decisions of multiple earlier arms. This means that the bias correction becomes conditional on multiple continuation events rather than a single interim outcome. In such complex cases, closed-form bias adjustments may no longer be tractable, and simulation-based estimation for the bias and variance may be required.

\newpage

\vspace*{2pc}
\noindent {\bf{Supplementary Material}}
Additional results from the simulation study. (pdf)

The GitHub repository (\url{https://github.com/pavlakrotka/NCC_InterimAnalysis}) 
contains the R code to reproduce the results of the simulation study.
	
\vspace*{1pc}
\noindent {\bf{Authors' Contributions}}
M.B.R., M.P. and P.K. conceived this research. P.K. prepared the initial draft of the text and performed the simulations. All authors discussed the results, provided comments and reviewed the manuscript.

\vspace*{1pc}

%%%%%%%%%%%%%%%%%%%%%%%%%%%%%%
%%%%%%%%%%%%%%%%%%%%%%%%%%%%%%
\noindent {\bf{Acknowledgments}}
This publication is supported by the predoctoral program AGAUR-FI ajuts (2024 FI-1 00401) Joan Oró, which is backed by the Secretariat of Universities and Research of the Department of Research and Universities of the Generalitat of Catalonia, as well as the European Social Plus Fund.

This work was supported by Grant PID2023-148033OB-C21 funded by MICIU/AEI/10.13039/501100011033 and by FEDER/UE; and the Departament d'Empresa i Coneixement de la Generalitat de Catalunya (Spain) under Grant 2021 SGR 01421 (GRBIO).

Marta Bofill Roig is a Serra Húnter Fellow. 

Martin Posch is a member of RealiseD, which is supported by the Innovative Health Initiative Joint Undertaking (IHIJU) under grant agreement No 101165912. The JU receives support from the European Union’s Horizon Europe research and innovation programme and COCIR, EFPIA, EuropaBío, MedTech Europe, and Vaccines Europe. Views and opinions expressed are those of the authors only. They do not necessarily reflect those of the Innovative Health Initiative Joint Undertaking and its members, EMA, the Norwegian Medical Products Agency or any pharmaceutical company.

%%%%%%%%%%%%%%%%%%%%%%%%%%%%%%

\vspace*{1pc}

\noindent {\bf{Conflict of Interest Statement}}
The authors declare no potential conflict of interest.

%%%%%%%%%%%%%%%%%%%%%%%%%%%%%%%%

%\bibliographystyle{biorefs}
\bibliography{references}

\newpage

%%%%%%%%%%%%%%%%%%%%%%%%%%%%%%%%

\appendix
\small

\section{Derivation of the bias of the model-based treatment effect estimator for Arm 2}\label{app_bias}

Let $\bar y_{ks}$ denote the sample mean in arm $k$ and period $s$. Assume patient $j$ enters the trial at time $t_j$, and is enrolled in period $s_j$. Further, assume a time trend defined by a function $f(t_j)$ of the patient entry time $t_j$ ($j=1, \ldots, N$), which is the same for all arms and additive on the model scale, as described in Bofill Roig et al. \cite{Bofill2022Model}. For each period $s$, define $m_s := E[f(t_j) \mid s_j=s]$ and assume that $E[f(t_j) \mid k_j=k, s_j=s] = m_s$ for all arms $k$ that are open for randomization in period $s$. We additionally assume that the variances of the outcomes $y_j$ are equal across periods. This approximately holds when the strength of the time trend is small relative to the response variance conditional on time.
Hence, the sample means in period 1 $\bar y_{01}$ and $\bar y_{11}$ follow normal distributions with means $\mu_0 + m_1$ and $\mu_1 + m_1$, respectively. In period 2, the sample means $\bar y_{02}$, $\bar y_{12}$, and $\bar y_{22}$ follow normal distributions with means $\mu_0 + m_2$, $\mu_1 + m_2$, and $\mu_2 + m_2$, respectively. 

Denote by $Z_{11} = (\bar y_{11} - \bar y_{01}) / \sigma \sqrt{1/n_{11} + 1/n_{01}}$ the Z-statistic from the interim analysis of Arm 1. The arm is dropped for futility if the p-value from the z-test is larger than a futility boundary $\alpha_F$ ($0 \le \alpha_F \le 1$), hence if $Z_{11} < \Phi^{-1}(1-\alpha_F) = c_F$. Equivalently, the arm is dropped for efficacy if the p-value from the z-test is smaller than an efficacy boundary $\alpha_E$ ($0 \le \alpha_E \le 1$, $\alpha_E < \alpha_F$), hence if $Z_{11} > \Phi^{-1}(1-\alpha_E) = c_E$.

The probability of Arm 1 to be stopped at the interim analysis is $\probP(Z_{11} < c_F)  + \probP(Z_{11} > c_E) = \Phi(c_F - \theta_1 / (\sigma\sqrt{1/n_{11} + 1/n_{01}})) + \Phi(\theta_1 / (\sigma\sqrt{1/n_{11} + 1/n_{01}}) - c_E) = \probP(stop)$. 

%To simplify the notation, define $\delta = c_1 - \theta_1 / (\sigma\sqrt{1/n_{11} + 1/n_{01}})$.

The marginal bias of $\Tilde{\theta}_2$ relative to $\theta_2$ is defined as:
\begin{align*}
    E \left[ \Tilde{\theta}_2 - \theta_2 \right] = E \left[ \Tilde{\theta}_2 \right] - \theta_2
\end{align*}
Based on the result of the interim analysis, the estimator $\Tilde{\theta}_2$ is either given by the separate analysis (if $Z_{11} < c_F$ or $Z_{11} > c_E$), or by the regression model (if $c_F \le Z_{11} \le c_E$). Let $I(\cdot)$ denote the indicator function. The expected value $E \left[ \Tilde{\theta}_2 \right]$ is then given by:

\begin{flalign*}
    & E \left[\Tilde{\theta}_2 \right] 
     = E \left[ \left[ \underbrace{\bar{y}_{22} - \bar{y}_{02}}_{(1)} \right] \cdot I\{Z_{11} < c_F \text{ or } Z_{11} > c_E\} \right]
    + E \left[ \left[\underbrace{\bar{y}_{22} - (1-\varrho) \bar{y}_{02}}_{(2)} - \underbrace{\varrho (\bar{y}_{01} + \bar{y}_{12} - \bar{y}_{11})}_{(3)} \right] \cdot I\{c_F \le Z_{11} \le c_E\} \right] &&
\end{flalign*}

\noindent Since the sample means from period 2 $\bar y_{02}$, $\bar y_{12}$, and $\bar y_{22}$ are independent of $Z_{11}$, and $\theta_2 = \mu_2 - \mu_0$ it follows:

\begin{flalign*}
    & \text{(i)} \ E[(1) \cdot I\{Z_{11} < c_F \text{ or } Z_{11} > c_E\}] = ((\mu_2 + m_2) - (\mu_0 + m_2)) \cdot \probP(stop) \\
    & \text{(ii)} \ E[(2) \cdot I\{c_F \le Z_{11} \le c_E\}] = (\mu_2 + m_2) (1-\probP(stop)) - (1- \varrho) (\mu_0 + m_2) (1-\probP(stop)) \\
    & \text{(iii)} \ E[(3) \cdot I\{ c_F \le Z_{11} \le c_E\}] = - \varrho (\mu_1 + m_2) (1-\probP(stop)) + \varrho (\mu_1 + m_1) (1-\probP(stop)) - \varrho (\mu_0 + m_1) (1 - \probP(stop)) + \\ 
    & \varrho \cdot \text{Cov} \left[ (\bar y_{11} - \bar y_{01}), I\{ c_F \le Z_{11} \le c_E\} \right] &&
\end{flalign*}

\noindent where in (iii) we have also used $E \left[ (\bar y_{11} - \bar y_{01}) \cdot I\{ c_F \le Z_{11} \le c_E\} \right] = E \left[ \bar y_{11} - \bar y_{01} \right] \cdot E \left[ I\{ c_F \le Z_{11} \le c_E\} \right] + \text{Cov} \left[ (\bar y_{11} - \bar y_{01}), I\{ c_F \le Z_{11} \le c_E\} \right]$.

\noindent The expression (i) + (ii) + (iii) simplifies to:

\begin{flalign*}
    E \left[\Tilde{\theta}_2 \right] = \theta_2 + \varrho \cdot \text{Cov} \left[ (\bar y_{11} - \bar y_{01}), I\{ c_F \le Z_{11} \le c_E\} \right] &&
\end{flalign*}

\noindent Note that $I (c_F \le Z_{11} \le c_E) = I \left( c_F \cdot \sigma \sqrt{\frac{1}{n_{11}} + \frac{1}{n_{01}}} \le \bar y_{11} - \bar y_{01} \le c_E \cdot \sigma \sqrt{\frac{1}{n_{11}} + \frac{1}{n_{01}}} \right)$. 
Moreover, note that the covariance $\text{Cov} \left[ X, I(a \le X \le b) \right] = \probP(a \le X \le b) (E[X | a \le X \le b] - E[X])$.

Hence, it follows:

\begin{flalign*}
    & E \left[\Tilde{\theta}_2 \right] = \theta_2 \ + \\ 
    & \varrho \cdot (1-\probP(stop)) \left(E \left[ \bar y_{11} - \bar y_{01} \mid c_F \cdot \sigma \sqrt{\frac{1}{n_{11}} + \frac{1}{n_{01}}} \le \bar y_{11} - \bar y_{01} \le c_E \cdot \sigma \sqrt{\frac{1}{n_{11}} + \frac{1}{n_{01}}} \right] - \theta_1 \right) &&
\end{flalign*}

\noindent Since $\bar y_{11} - \bar y_{01} \sim \mathcal{N} \left( \theta_1, \sigma^2 \left( \frac{1}{n_{01}} + \frac{1}{n_{11}} \right) \right)$, we can use the following formula for the expectation of the truncated normal distribution: Let $X \sim N(\mu, \sigma^2)$, then $E[ X \mid a < X < b] = \mu - \sigma \frac{\phi((b - \mu) / \sigma) - \phi((a - \mu) / \sigma)}{\Phi((b - \mu) / \sigma) - \Phi((a - \mu) / \sigma)}$. Therefore, it follows: 

\begin{flalign*}
    & E \left[\Tilde{\theta}_2 \right] = \theta_2 \ + \\
    & \varrho \cdot \left(1 - \probP(stop)\right) \left( \sigma \sqrt{\frac{1}{n_{11}} + \frac{1}{n_{01}}} \cdot \frac{\phi \left(c_F - \frac{\theta_1}{\sigma \sqrt{\frac{1}{n_{11}} + \frac{1}{n_{01}}}} \right) - \phi \left(c_E - \frac{\theta_1}{\sigma \sqrt{\frac{1}{n_{11}} + \frac{1}{n_{01}}}} \right)}{\Phi \left(c_E - \frac{\theta_1}{\sigma \sqrt{\frac{1}{n_{11}} + \frac{1}{n_{01}}}} \right) - \Phi \left(c_F - \frac{\theta_1}{\sigma \sqrt{\frac{1}{n_{11}} + \frac{1}{n_{01}}}} \right)} \right) &&
\end{flalign*}

\noindent Hence, the bias of $\Tilde{\theta}_2$ relative to $\theta_2$ is given by:

\begin{align*}
    E \left[ \Tilde{\theta}_2 - \theta_2 \right] = \varrho \cdot \sigma \sqrt{\frac{1}{n_{11}} + \frac{1}{n_{01}}} \cdot \left( \phi \left(c_F - \frac{\theta_1}{\sigma \sqrt{\frac{1}{n_{11}} + \frac{1}{n_{01}}}} \right) - \phi \left(c_E - \frac{\theta_1}{\sigma \sqrt{\frac{1}{n_{11}} + \frac{1}{n_{01}}}} \right) \right)
\end{align*}

\noindent The expression for the conditional bias in equation \eqref{eq_bias_cond} follows from the definition of the conditional expectation of a continuous random variable X conditional on an event A: 

\begin{align*}
    E[X \mid A] = \frac{E[X \cdot I(A)]}{\probP(A)}
\end{align*}

\noindent Thus, the expectation of $\tilde{\theta}_2$ conditional on $c_F \le Z_{11} \le c_E$ is defined as follows:

\begin{flalign*}
    E[\tilde{\theta}_2 \mid c_F \le Z_{11} \le c_E] = \frac{E \left[ \left[\bar{y}_{22} - (1-\varrho) \bar{y}_{02} - \varrho (\bar{y}_{01} + \bar{y}_{12} - \bar{y}_{11}) \right] \cdot I\{ c_F \le Z_{11} \le c_E\} \right]}{\probP(c_F \le Z_{11} \le c_E)} &&
\end{flalign*}

\noindent Through analogous calculations as we used for the unconditional expectation, we get the following result for the conditional bias:

\begin{align*}
    E \left[ \Tilde{\theta}_2 - \theta_2 \mid c_F \le Z_{11} \le c_E \right] = \varrho \cdot \sigma \sqrt{\frac{1}{n_{11}} + \frac{1}{n_{01}}} \cdot \frac{\phi \left(c_F - \frac{\theta_1}{\sigma \sqrt{\frac{1}{n_{11}} + \frac{1}{n_{01}}}} \right) - \phi \left(c_E - \frac{\theta_1}{\sigma \sqrt{\frac{1}{n_{11}} + \frac{1}{n_{01}}}} \right)}{\Phi \left(c_E - \frac{\theta_1}{\sigma \sqrt{\frac{1}{n_{11}} + \frac{1}{n_{01}}}} \right) - \Phi \left(c_F - \frac{\theta_1}{\sigma \sqrt{\frac{1}{n_{11}} + \frac{1}{n_{01}}}} \right)}
\end{align*}

\noindent Equation \eqref{eq_bias_cond_null} follows from substituting 0 for the treatment effect $\theta_1$:

$$
 E \left[ \Tilde{\theta}_2 - \theta_2 \mid c_F \le Z_{11} \le c_E \right]_{\theta_1 = 0}  = \varrho \cdot \sigma \sqrt{\frac{1}{n_{11}} + \frac{1}{n_{01}}} \cdot \frac{\phi \left( c_F \right) - \phi \left( c_E \right)}{\alpha_F - \alpha_E} 
$$

%%%%%%%%%%%%%%%%%%%%%%%%%%%%%%%%
%Figures and tables
%...
%Supplementary material (file in the tex folder)

%\includepdf[pages=-]{Supp\_material.pdf}
\includepdf[pages=-]{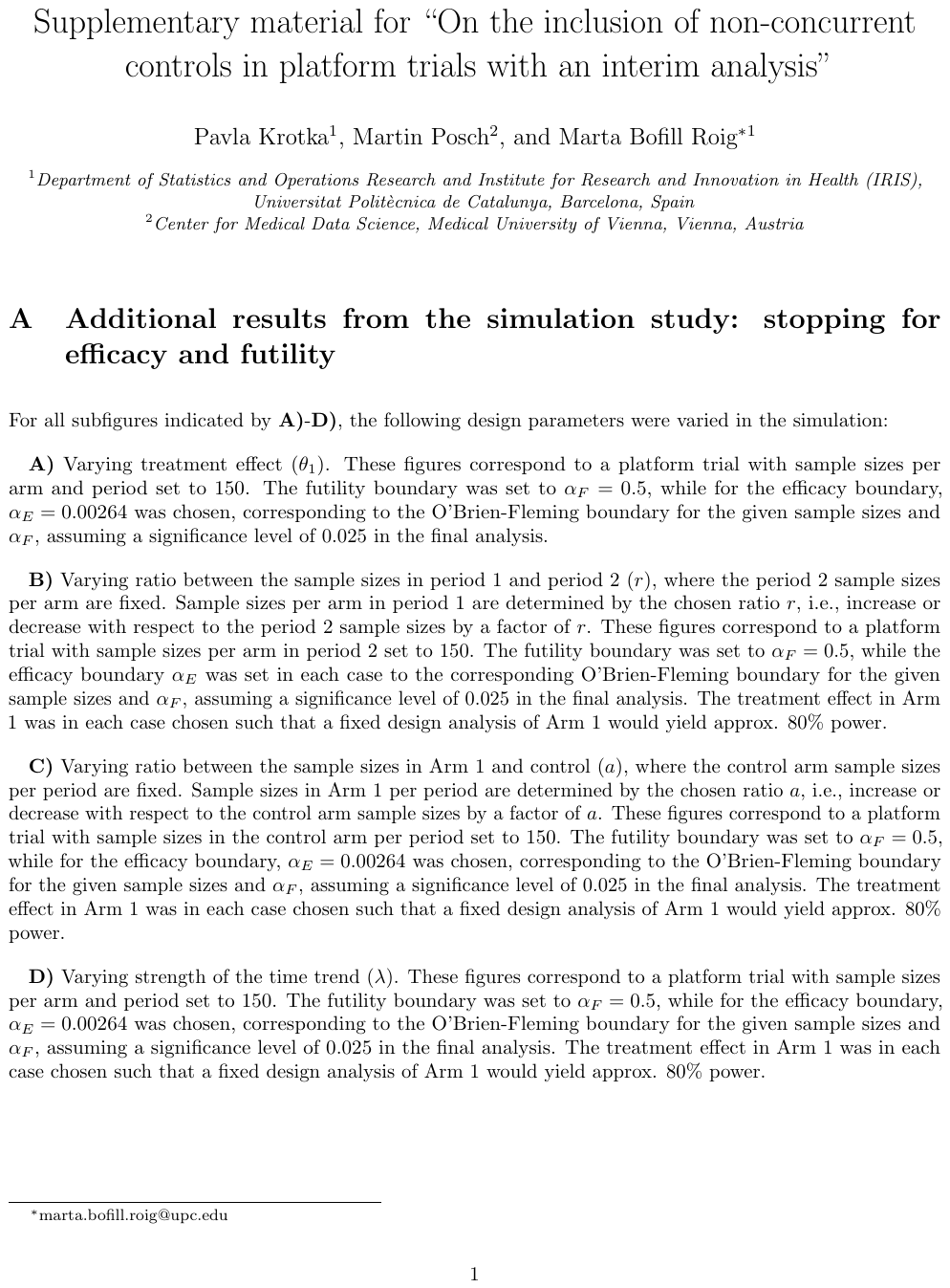}

\clearpage

\end{document}